\documentstyle[12pt,epsf]{article}
{}{}
{}{}
\begin{document}                

\begin{minipage}{14cm}
\hspace*{8.5cm} {\bf Preprint JINR}\\
\hspace*{9.cm}{\bf E2-2003-200}\\
\hspace*{9.0cm}{\bf Dubna, 2003}\\[0.4cm]

\end{minipage}

\vspace*{2.cm}

\begin{center}
{\large \bf Z-scaling from tens of GeV to TeV}
\vskip 1.cm {\bf M. V. Tokarev\footnote{E-mail: tokarev@sunhe.jinr.ru}}
\vskip 0.5cm
{\it
Veksler and Baldin Laboratory of High Energies,\\
Joint Institute for Nuclear Research,\\
 Dubna,  Russia
}

\vskip 1cm
\begin{abstract}
The concept of $z$-scaling reflecting the general features of
internal particle substructure, constituent interaction and
mechanism of particle formation at high-$p_T$ is reviewed.
 Experimental
data on inclusive cross sections obtained at U70, ISR, SpS, RHIC and
Tevatron are used in the analysis. The properties of data
$z$-presentation such as the energy independence, power
law and $A$-dependence are discussed.
The properties  of $z$-scaling are argued to be connected with the fundamental
symmetries such as self-similarity, locality and fractality. The use of $z$-scaling
to search for new physics phenomena in hadron-hadron and
hadron-nucleus collisions is suggested.
RHIC data used in our new analysis  confirm $z$-scaling in $pp$ collisions.
High-$p_T$ particle spectra at LHC energies are
predicted.  Violation of $z$-scaling characterized by the
change of  anomalous fractal dimension is considered as a new and
complementary signature of new physics phenomena.
\end{abstract}

\vskip 1cm

{\it
 Presented at the 7th International Workshop\\
 "Relativistic Nuclear Physics: from Hundreds of MeV to TeV",\\
 August 25-30, 2003, Stara Lesna, Slovak Republic}

\end{center}

\newpage

{\section{Introduction}}

More prominent properties of particle are observed at high energy
and transverse momentum
\cite{Bjorken}-\cite{Brodsky}.
The kinematic regime is used to
perform  calculations of physical quantities in the framework of
perturbative  Quantum Chromodynamics (QCD). Deviations of theoretical
results obtained in a high-$p_T$ region using available experimental data
are often considered  as manifestations of new physics phenomena.
At the same time we should note that nonperturbative effects
are not well controlled by the theory. Therefore search for
new scaling features of particle interaction in the region
is of interest for development of the theory.

The fundamental problem of high energy physics is the origin of mass,
spin and charge of particles. The study of particle interactions over a wide
kinematic range and especially at small scales is necessary to
resolve the problem  and  understand underlying physics phenomena.
One assumes that all types of interactions are unified at small
scales.
New ideas  and theories such as extra dimensions, anisotropy and
fractality of space-time, quark compositeness, theories of Grand Unification,
Super Symmetry and Super Gravity are intensively developed.

New feature ($z$-scaling) of high-$p_T$ particle production in hadron-hadron
and hadron-nucleus collisions at high energies was established in
\cite{Z96}-\cite{Rog2}.
The scaling function $\psi$ and scaling variable $z$ are expressed via
experimental quantities such as the inclusive cross section
$Ed^3\sigma/dp^3$ and the multiplicity density of charged particles
$dN/d\eta$.
Data $z$-presentation is found to reveal symmetry
properties (energy independence, A-dependence, power law).
The properties  of $\psi$ at high $z$ are assumed  to be
relevant to the structure of space-time at small scales \cite{Zbor,Nottale}.
The function  $\psi(z)$ is interpreted as the
probability density to produce a particle with a formation length $z$.
The concept of $z$-scaling and the method of data analysis are
developed for description  of different particles
(charged \cite{Z99,Z01,Toivonen} and neutral \cite{Rog1,Rog2} hadrons,
direct photons \cite{Potreben,Efimov}, jets \cite{Dedovich})
 produced in high energy
hadron-hadron and hadron-nucleus interactions.
 The proposed method is complementary to a method of direct
calculations developed in the framework of QCD \cite{QCD}
and methods based on Monte Carlo generators \cite{JETRAD}-\cite{NEXUS}.
Therefore we consider that the use
of the method allow to reduce some theoretical uncertainties
which are ambiguously estimated by theory.

In the report the general concept of $z$-scaling, the properties
of data $z$-presentation and some new results of data analysis
are presented. The fundamental principles such as
self-similarity, locality, fractality and scale-relativity are
formulated and discussed in the framework of the $z$-scaling concept.
Verification $z$-scaling validity at RHIC and LHC is suggested.
Violation $z$-scaling is considered to be indication of new
physics phenomena.

\vskip 0.5cm
{\section{Z-scaling}}

In the section we discuss underlying ideas of $z$-scaling.
A general scheme of data $z$-presentation is described.
Physical meaning of the introduced quantities is explained.

\vskip 0.5cm
{\subsection{Locality}

The idea of $z$-scaling is based on the assumptions \cite{Stavinsky}
that gross feature of inclusive particle distribution of the
process (\ref{eq:r1}) at high energies can be described in terms
of the corresponding kinematic characteristics
\begin{equation}
M_{1}+M_{2} \rightarrow m_1 + X
\label{eq:r1}
\end{equation}
of the constituent subprocess  written in the symbolic form (\ref{eq:r2})
\begin{equation}
(x_{1}M_{1}) + (x_{2}M_{2}) \rightarrow m_{1} +
(x_{1}M_{1}+x_{2}M_{2} + m_{2})
\label{eq:r2}
\end{equation}
satisfying the condition
\begin{equation}
(x_1P_1+x_2P_2-p)^2 =(x_1M_1+x_2M_2+m_2)^2.
\label{eq:r3}
\end{equation}
The equation is the expression of locality of  hadron interaction at
constituent level. The $x_1$ and $x_2$ are fractions of the incoming
momenta $P_1$ and $P_2$  of  the colliding objects with the masses $M_1$
and $M_2$. They determine the minimum energy, which
is necessary for production of the secondary particle with
the mass $m_1$ and the four-momentum $p$.
The parameter $m_2$ is introduced to satisfy the internal
conservation laws (for baryon number, isospin, strangeness, and so on).

The equation (\ref{eq:r3}) reflects minimum recoil mass hypothesis in the
elementary subprocess.
To connect kinematic and structural
characteristics of the interaction, the quantity
$\Omega$ is introduced. It is chosen in the form
\begin{equation}
\Omega(x_1,x_2) = m(1-x_{1})^{\delta_1}(1-x_{2})^{\delta_2},
\label{eq:r5}
\end{equation}
where $m$ is a mass constant and $\delta_1$ and $\delta_2$
are factors relating to the anomalous fractal dimensions  of
the colliding objects. The fractions $x_{1}$ and
$x_{2}$  are determined  to maximize the value of $\Omega(x_1,x_2)$,
simultaneously fulfilling the condition (\ref{eq:r3})
\begin{equation}
{d\Omega(x_1,x_2)/ dx_1}|_{x_2=x_2(x_1)} = 0.
\label{eq:r6}
\end{equation}
The fractions  $x_{1}$  and $x_2$ are equal to unity along
the phase space limit and
cover the full phase space accessible at any energy.

\vskip 0.5cm
\subsection{Self-similarity}}

Self-similarity is a scale-invariant property connected with dropping of
certain dimensional quantities out of physical picture of the interactions.
It means that dimensionless quantities for the description of physical processes
are  used.
The scaling function
$\psi(z)$ depends in a self-similar manner on a single dimensionless variable
$z$. It is expressed  via the invariant cross section
$Ed^3\sigma/dp^3$ as follows

\begin{equation}
\psi(z) = -{  { \pi s} \over { ({dN/d\eta}) \sigma_{in}} } J^{-1} E{ {d^3\sigma} \over {dp^3}  }
\label{eq:r7}
\end{equation}
Here, $s$ is the center-of-mass collision energy squared, $\sigma_{in}$ is the
inelastic cross section, $J$ is the corresponding Jacobian.
The factor $J$ is the known function of the
kinematic variables, the momenta and masses of the colliding and produced particles.

The function $\psi(z)$ is normalized as follows
\begin{equation}
\int_{0}^{\infty} \psi(z) dz = 1.
\label{eq:r8}
\end{equation}
The relation allows us to interpret the function $\psi(z)$
as a probability density to produce
a particle with the corresponding value of the variable $z$.

 We would like to emphasize that  the existence of the function  $\psi(z)$
 depending on a single dimensionless  variable $z$ and
 revealing scaling properties is not evident in advance.
 Therefore the method proposed to construct  $\psi(z)$ and $z$
 could be only proved  a posteriori.

\vskip 0.5cm
{\subsection{Fractality}}

Principle of fractality states that variables used in the
description of the process diverge in terms of the resolution.
This property is characteristic for the scaling variable
\begin{equation}
z = z_0 \Omega^{-1},
\label{eq:r9}
\end{equation}
where
\begin{equation}
z_0 = \sqrt{ \hat s_{\bot}} / (dN/d\eta).
\label{eq:r10}
\end{equation}
The variable $z$ has character of a fractal measure.
For the given production process (\ref{eq:r1}),
its finite part $z_0$ is the ratio
of the transverse energy released in the
binary collision of constituents (\ref{eq:r2})
and the average multiplicity density $dN/d\eta|_{\eta=0}$.
The divergent part
$\Omega^{-1}$ describes the resolution at which the collision of
the constituents can be singled out of this process.
The $\Omega(x_1,x_2)$ represents relative number of all initial
configurations containing the constituents which carry fractions
$x_1$ and $x_2$ of the incoming momenta.
The $\delta_1$ and $\delta_2$ are the anomalous fractal
dimensions of the colliding objects (hadrons or nuclei).
The momentum fractions $x_1$ and $x_2$ are determined in a way to
minimize the resolution $\Omega^{-1}(x_1,x_2)$ of the fractal
measure $z$ with respect to all possible sub-processes
(\ref{eq:r2}) subjected to the condition (\ref{eq:r3}).
The variable $z$ was interpreted  as a particle formation length.

  As we will show later the scaling function of high-$p_T$ particle production
  is described  by the power law, $\psi(z) \sim z^{-\beta} $.
  Both quantities, $\psi$ and $z$, are scale dependent.
  Therefore we consider the high energy
  hadron-hadron interactions as interactions of fractals. In the
  asymptotic  region the internal structure of particles, interactions of their constituents and
   mechanism of real particle formation manifest self-similarity over a
   wide scale range.

\vskip 5mm
{\subsection{Scale-relativity}}

 The properties of particle interactions in the space-time reflect
 symmetries of Nature.
 The principle of motion relativity  has been used to formulate
 non-relativistic and relativistic theory.
 The special theory deals with only inertial coordinate systems
 while the expression of physical laws in the general theory of relativity
 should be written into any curvilinear coordinate system.
 Principle of general relativity states that
 "the laws of physics must be of such a nature that they apply to systems
 of references in any kind of motion".
 Application of relativity principle can be extended to state of scale
 of reference system \cite{Nottale}.
There are convincing evidence to consider that scale as well as
other quantities  characterizing a reference frame should
be used to describe a particle state in a high-$p_T$ range.
In this range elementary probes such as direct photons,
high-$p_T$ hadrons and jets are not point-like objects.
They have a complicated structure. The last is resulted from
interactions of quarks, gluons and heavy bosons
which are fundamental objects of theory.
Therefore experimentally measurable quantities should depend on a ratio between
scales of a studied object and a probe. In other words
the variables used in the description of the process depends
on the resolution.

A generalization of the motion-relativity principle to the scale-relativity  principle
requires that "the laws of physics must be of such a nature that they apply
to systems of references in any kind of motion and whatever its state of scale".

In the framework of data $z$-presentation the mechanism of particle formation
is described by the scaling function $\psi(z) $. Both quantities, $\psi$ and $z$,
depend on the resolution $\Omega^{-1}$  while the anomalous fractal dimension $\delta$
as found from our analysis of numerous experimental data to be resolution independent.
We consider that the study of $z$-scaling of high-$p_T$ particle production
over a wide kinematic range of $p_T$ and $\sqrt s$ and determination of $\delta$
could give new insight in theory of scale-relativity \cite{Zbor,Nottale}.

\vskip 0.5cm
{\section{Properties of data $z$-presentation}}

In the section we present and discuss some properties of data $p_T$-
and $z$-presentation of high-$p_T$ particle production in $pp, \bar pp$
and $pA$ collisions. We  show that the scaling functions for different
processes reveal the same properties. They are the energy independence of $\psi$,
the power behavior of $\psi$ at high-$z$ and $A$-dependence.

\vskip 5mm
{\subsection{Energy independence of $\psi(z)$}}

It is well known that numerous experimental data on high-$p_T$ particle
spectra manifest the strong dependence  on collision
energy $\sqrt s$. The effect enhances  as the transverse momentum  of produced
particle increases.

\vskip 5mm
{\subsubsection{$\pi^+$-mesons }}

Figures 1(a) and 2(a) show the dependence of the inclusive cross sections
of $\pi^+$  and $K^+$ meson  production in $pp$  collisions on the transverse momentum
$p_T$ at incident proton momentum  $p_L = 70, 200, 300, 400$ and  800 GeV/c and an angle
$\theta_{cm}\simeq 90^o$.  The data were obtained at Protvino \cite{Protvino}
and Batavia \cite{Cronin,Jaffe}.
Transverse momenta of produced particles shown in Figures 1(a) and 2(a)
change from 1  to 10 GeV/c. We would note that data \cite{Cronin}
and \cite{Jaffe} corresponding to the momentum $p_L=400$ GeV/c are complementary
and are in a good agreement each other. As seen from Figures 1(a) and 2(a)
particle spectra have power behavior at $p_L = (200-800)$ GeV/c.
The effect of kinematic boundary is visible at the end of spectrum
 at $p_L =70$ GeV/c.

 The energy independence of data $z$-presentation means that
 the scaling function $\psi(z)$ has the same shape for different
 $\sqrt s$ over a wide $p_T$ range.

 As seen from  Figures 1(b) and 2(b)  $z$-presentation of the same data sets
 demonstrates the energy  independence  of $\psi(z) $  over a wide collision energy and
 transverse momentum  range. We would like to emphasize that the
 data \cite{Jaffe} used
 in our new analysis confirm our earlier results \cite{Z96,Varna}.

\vskip 5mm
{\subsubsection{$\pi^0$-mesons}}

  The PHENIX Collaboration published the new data \cite{Phenix} on inclusive spectrum
  of $\pi^0$-mesons produced in $pp$ collisions in central rapidity range
  at RHIC energy $\sqrt s = 200$ GeV.  The transverse momenta of $\pi^0$-mesons
  were measured up to 13 GeV/c.

 The data $p_T$- and $z$-presentations for $\pi^0$-meson spectra obtained at ISR
 \cite{Angel}-\cite{Eggert}
 and RHIC \cite{Phenix}  are shown in Figures 3(a) and 3(b).  One can see that $p_T$-spectra
 of $\pi^0$-meson production reveal the properties similar to that found for charged hadrons.
 The new data \cite{Phenix} on $\pi^0$-meson inclusive cross sections obtained at RHIC
 as seen from Figure 3(b) are in a good agreement with our earlier results \cite{Rog1}.
 Thus we can conclude that available experimental data on high-$p_T$  $\pi^0$-meson production
 in $ pp$ collisions confirm the property of the energy independence of $\psi(z)$
 in $z$-presentation.

\vskip 5mm
{\subsubsection{Charged hadrons}}

 The STAR Collaboration published new data  \cite{STAR} on inclusive cross
 sections of charged hadrons produced
 in $pp$  collisions at RHIC energy $\sqrt s = 200$ GeV.
  The RHIC data and other ones obtained at U70 \cite{Protvino}, ISR \cite{Alper} and Tevatron
  \cite{Cronin,Jaffe} are shown in Figure 4(a).
  Charged hadron spectra were measured  over a wide  kinematic  range  $\sqrt s = (11.5-200)$ GeV and
  $p_T = (0.5-9.5) $ GeV/c.  The strong energy dependence and the power behavior
  of particle $p_T$-spectrum are found to be clearly. The energy independence of
  data $z$-presentation shown in Figure 4(b) is confirmed.
  It is of interest to verify the asymptotic behavior of $\psi$ at $\sqrt s =200$ GeV
  and reach value of $z$ up to 30 and more.

\vskip 5mm
{\subsubsection{Direct ${\gamma}$}}

   Direct photons are considered as the best probes of constituent
   interactions at high-$p_T$. The calculations of direct photon cross sections
   in the framework of QCD are developed in next-to-leading order \cite{Aurenche}.
   The basic mechanisms
   of direct photon production are considered to be  Compton scattering ($gq\rightarrow \gamma q$),
   and annihilation process ($\bar qq \rightarrow \gamma g$).  These are direct mechanisms of photon
   production.  In high-$p_T$ range  direct photons are also produced indirectly via bremsstrahlung
   of quarks   ($qq\rightarrow qq\gamma,\   qg\rightarrow qg\gamma $).
    The contribution of indirect mechanisms of
   high-$p_T$ photon production can be significant. However there are significant theoretical
   uncertainties  due to the choice of structure and fragmentation functions and renormalization,
   factorization and fragmentation scales. Therefore any reliable estimates of direct photon
   cross sections  allowing to reduce some of theoretical uncertainties  is of interest.

   The results of data analysis for direct photon production in $\bar pp$ collisions
   are presented in Figure 5. Inclusive cross sections versus the transverse momentum
   at $\sqrt s = (24-1800)$ GeV over the range $p_T = (4-110)$ GeV/c
   are shown in Figure 5(a). Data used in the analysis are obtained by UA1 \cite{UA1},
    UA2 \cite{UA2}, UA6 \cite{UA6p},
   CDF \cite{CDFpho}  and D0 \cite{D0pho} Collaborations.
    Data $z$-presentation (see Figure 5(b)) demonstrates
   the energy independence of the scaling function $\psi$ of high-$p_T$ direct photon
   production over a wide kinematic range.

\vskip 5mm
{\subsubsection{Jets}}

   First observation of jets in $\bar pp$ collisions at SpS  was considered
   as compelling confirmation of parton structure of hadrons.
   In high energy collisions of hadrons copiously jet production
   due to hard parton scattering
   is observed. A jet represents a group of moving collimated particles.
   In the framework of QCD jets are distinguished to can be quark and gluon.
   Quark  and gluon  jets are initiated by fastest quark and gluon, respectively.
   It should be noted that the mechanism of jet formation is insufficient explored
   and not clearly  understood till now.
   Therefore we hope that scaling features of jet production
   could be useful to obtain
   additional constraints for models of jet formation.

   In Figures 6(a) we show the invariant cross sections of inclusive jet production
   in $\bar pp$ collisions at $\sqrt s = 630$ and 1800 GeV. These experimental
   data  are obtained by CDF \cite{CDFjet} and D0 \cite{D0jet} Collaborations.
   A clear energy dependence of the cross section is observed to be.
   Difference between cross sections at $\sqrt s = 630$ and  1800 GeV
   increases with transverse energy of jet.  Data $z$-presentation shown in Figure 6(b)
   demonstrates independence on collision energy $\sqrt s$.
   The anomalous fractal dimensions for jet production in $pp$ and $\bar pp$
   collisions was found to be constant and equal to 1 \cite{Dedovich}.

\vskip 0.5cm
{\subsection{A-dependence of $\psi(z)$}}

 A comparison of particle yields in hadron-hadron and hadron-nucleus collisions is
a basic method to study the nuclear matter influence on particle production.
The elementary process is considered as probe of more complex system like nucleus.
The difference  between  cross sections of particle production
on free and bound nucleons was considered as an indication of unusual physics
phenomena like EMC-effect, $J/\psi$-suppression and Cronin effect.

A change of the shape of  $p_T$ spectra is considered to be evidence that
the mechanism of particle formation  in nuclear matter  is modified.
Therefore it is convenient to compare scaling functions corresponding to
different $pA$ processes over a wide range of $\sqrt s $  and $p_T$.

The search for scaling features of particle formation in $pA$ as well as in $pp$ collisions
and  the study of their dependence on the atomic weight $A$ is of interest for development of theory.

$A$-dependence of $z$-scaling of hadron production in $pA$ collisions
was studied in \cite{Z01,Rog2}.
It was established $z$-scaling for different nuclei ($A=D-Pb$) and types of produced
particles ($\pi^{\pm,0}, K^{\pm},
\bar p$). To compare the scaling functions for different nuclei
the symmetry transformation
\begin{equation}
z\rightarrow \alpha_A z,\ \ \  \psi \rightarrow \alpha_A^{-1} \psi
\label{eq:r11}
\end{equation}
has been used. The parameter $\alpha$ of the scale transformation (\ref{eq:r11}) depends
on the atomic weight $A$. It was parameterized by the formula
$\alpha(A)=0.9A^{0.15}$.

Figures 7(a) demonstrates  the spectra of $\pi^+$-mesons produced in proton-nucleus
collisions at $\sqrt s =11.5$ and 27.4 GeV and $\theta_{cm}^{NN}\simeq 90^0$.
Our new data analysis includes experimental data obtained at Protvino \cite{Protvino}
and Batavia \cite{Cronin,Jaffe}. The data \cite{Jaffe} extend transverse momentum range
up to $p_T = 8.5$ GeV/c. A good compatibility  of \cite{Cronin} and \cite{Jaffe} data sets
in the overlapping region was observed.
Solid and dashed lines are obtained by fitting  of the data
for $W, Pb$ and $D$, respectively. They demonstrate
the strong dependence of $p_T$-spectra on collision energy $\sqrt s $.

Figure 7(b) shows the $z$-presentations of the same data.
The obtained results is the new confirmation of $z$-scaling of  high-$p_T$ hadron
production in $pA$ collisions.
The universality of the scaling function $\psi$ for different nuclei means that mechanism of
high-$p_T$ particle formation in nuclear matter reveals property of self-similarity.

\vskip 0.5cm
{\subsection{Power law}}

One of the general properties of data $z$-presentation is the power
law of scaling function
\begin{equation}
\psi(z) \sim z^{-\beta}.
\label{eq:r12}
\end{equation}
Such behavior of $\psi$ as seen from Figures 1(b)-7(b)
is observed for different particles (hadrons, direct photons and jets)
produced at $z>4$. The data sets demonstrate a linear $z$-dependence of
$\psi(z)$ on the log-log scale at high $z$.
The quantity  $\beta $ is the slope parameter.
The value of the slope parameter $\beta$ is found to be constant with high accuracy
and it is independent of energy $\sqrt s$ over a wide high transverse momentum   range.
Some indications (for $\pi^0$ mesons, charged hadrons, direct photons, jets) are obtained
that the value of slope parameter for $pp$ is larger than the corresponding  value
for $\bar pp$ collisions, $\beta_{pp} > \beta_{\bar pp}$.

The existence of the power law, means, from our point of view, that
the mechanism of particle formation reveals fractal behavior.

\vskip 0.5cm
{\section{Multiplicity charged particle density}}

The important ingredient of $z$-scaling concept is the multiplicity
density of charged particles, $dN/d\eta (s,\eta)$.
The scaling function $\psi$  and the scaling variable $z$
is proportional to $[dN/d\eta ]^{-1}$.
In the first case the quantity is included in the expression (\ref{eq:r6})
to normalize the function $\psi $ and to give the physical meaning for it as
a probability density to produce a particle with formation length $z$.
In the second one (\ref{eq:r8}) the multiplicity density is taken at $\eta = 0$.
Therefore $z$ is proportional to energy of elementary subprocess per one
particle produced in the initial hadron collision.

The energy dependence of the multiplicity charged particle density for
inelastic and non-single diffractive $pp$ and $\bar pp$ collisions  is shown
in Figure 8(a) \cite{Thome}.
The collision energy $\sqrt s $ changes from $14$  to 1800 GeV.
New data for $dN/d\eta (s,\eta)$ as well as for inclusive cross section
$Ed^3\sigma/dp^3$  for $pp$ collisions at RHIC energies are of interest
for verification of $z$-scaling.

\vskip 0.5cm
{\section{$\gamma/\pi^0$ ratio for $pp$ and $\bar pp$}}

  The properties of the scaling function for direct
 $\gamma$ and $\pi^0$-meson  can be used to estimate the dependence
 of the $\gamma / \pi^0$ ratio of inclusive cross sections on
 transverse momentum at LHC energies.

 The asymptotic behavior of $\psi(z)$ was found to be described by the  power law for
 $\pi^0$-meson and direct photon production in $pp$ and $\bar pp$ collisions.
 The slope parameters are satisfied to the relations
 $\beta_{pp}^{\gamma} > \beta_{\bar pp}^{\gamma}$,
  $\beta_{pp}^{\pi^0} >  \beta_{\bar pp}^{\pi^0}$,
 $\beta_{pp}^{\pi^0} > \beta_{pp}^{\gamma}$ and
 $\beta_{\bar pp}^{\pi^0} > \beta_{\bar pp}^{\gamma}$.

 As seen from  Figure 3(b) the cross section data \cite{Phenix}
 of $\pi^0$-meson production in $pp$ collisions
 obtained by  the PHENIX collaboration at RHIC are in a good agreement with
 the asymptotic behavior of $\psi(z)$.

 Figure 8(b) shows the $\gamma/\pi^0$ ratio of inclusive cross sections
 as a function of the transverse momentum $p_T$  at $\sqrt s = 5.5$ and  14. TeV.
 The ratio was found to be different for $pp$ and $\bar pp$ collisions. It increases with $p_T$.
 The ratio has the cross-over point at $p_T\simeq (60-70)$ GeV/c  and
  $p_T \simeq  (110-130)$ GeV/c for $pp$ and $\bar pp$ collisions, respectively.

\vskip 0.5cm
{\section{$z-p_T$ plot}}

The $z-p_T$ plot is the dependence of the variable $z$
on the transverse momentum  $p_T$ of produced
particle  for a given process.
The plot allows us to determine the high transverse momentum
range interesting for verification of $z$-scaling  and
experimentally inaccessible up to now.

 Figure 9(a) shows the  $z-p_{T}$ plot for the $pp\rightarrow \pi^+ X$
 process  at $\sqrt s = (24-14000)$ GeV.
 As seen from Figure 1(b) the scaling function $\psi(z)$
 was measured up to $z\simeq 30$.
 The function $\psi(z)$ demonstrates the power behavior at $z>4$.
 Therefore the kinematic range $ z > 30$ is of more
 preferable for experimental investigations of
 $z$-scaling violation. The boundary $z=30$ corresponds to the different
 values of the transfers momentum $p_T$ depending on collision energy $\sqrt s$.

 Figure 9 (b) shows our predictions of the dependence of the inclusive
 cross section  $Ed^3\sigma /dp^3$ on the transverse momentum  $p_T$
 for $\pi^+$-mesons  produced  in $pp$ collisions  at the
 ISR, RHIC and LHC energies and an angle  $\theta_{cm} = 90^0$.
 The verification of the predictions is of interest
 to determine the region of the scaling validity and search for new physics phenomena.

\vskip 0.5cm
{\section{ "$\delta$-jump" }}

Mechanism of particle formation at high transverse momenta in
$z$-presentation is described by the power law (\ref{eq:r12}).
Such behavior depends on the values of the anomalous fractal
dimension of colliding particles, $\delta_1$ and $\delta_2$.
The dimensions for hadrons, direct photons and jets produced in
$pp$ collisions  were found  to satisfy the relation
$\delta_h < \delta_{\gamma} < \delta_{jet}$ and to be independent of $\sqrt s $ and $p_T$.
The anomalous fractal dimension for nucleus $\delta_A$
is expressed via the dimension for
nucleon  $\delta_N$ as follows $\delta_A = A\cdot \delta_N$.

Figure 10(a) shows the dependence of the  anomalous fractal
dimension $\delta$ for $\pi^0$-meson production in $pp$ and $\bar pp$ \cite{Banner}
 collisions on energy $\sqrt s$.  The value of $\delta_h=0.5$ used
in our previous data analysis
is confirmed by the new data \cite{Phenix} on inclusive cross section  (see Figure 3(b))
obtained by the PHENIX Collaboration at RHIC. Figure 10(b) gives
evidence that values of the slope parameter $\beta $ of the scaling function
for $pp$ and $\bar pp$ collisions differ each other at $z>6$.

The change of the fractal dimension $\delta$  or "$\delta$-jump"
is considered as an indication on new mechanism of particle formation.
It is assumed that the energy dependence of the quantity is especially
sensitive in the high-$p_T$ range. Therefore the study of $z$-scaling at
higher $\sqrt s$ and $p_T$ is of interest for search for new physics phenomena.


\vskip 0.5cm
{\section{ Direct $\gamma$ and $\eta^0$-meson yields in $pp$ and  $pPb$ \\
\hspace*{1.5cm} collisions at RHIC and LHC}}

The scaling properties of data $z$-presentation  for $pp$ and $pA$ collisions
can be used to estimate particle yields in the kinematic region experimentally
inaccessible at present time  and to compare with other model predictions.

Figures 11 and 12 shows our predictions of the dependence of the inclusive
cross section  $Ed^3\sigma /dp^3$ on the transverse momentum  $p_T$
for direct photon (a) and $\eta^0$-mesons (b) produced
 in $pp$  and $pPb $ collisions  at RHIC and LHC energies and an angle
$\theta_{cm}^{NN} = 90^0 $.
 The data on the cross sections \cite{Kou2,UA6p,R806}
 obtained at ISR energy
  $\sqrt s = (24-63)$ GeV are also shown for comparison.

\vskip 5mm
{\section{Conclusions}}

The general concept of $z$-scaling for particle production
in hadron-hadron and hadron-nucleus collisions
with high transverse momenta  was reviewed.
The development of new method of data analysis based on data $z$-presentation
was presented.
The scaling function $\psi(z)$ and scaling variable $z$ were shown
to be expressed via the experimental quantities, the invariant
inclusive cross section
$Ed^3\sigma/dp^3$  and the multiplicity  charged  particles  density
$dN/d\eta (s,\eta)$.

Physical interpretation of the scaling function $\psi(z)$ and variable $z$
as a probability density to produce a particle with  formation length $z$
was  argued.
The quantity  $z$ was shown to reveal the property of the fractal measure
and $\delta_{1,2} $ are the anomalous fractal dimensions of colliding
particles.
It was argued that $z$-scaling  reflects the fundamental symmetries
such as locality, self-similarity,  fractality  and scale relativity.

Results of new analysis of experimental data on the inclusive cross
sections obtained at U70, ISR, SpS, Tevatron  and RHIC were presented.
The scaling properties of data $z$-presentation such as
the energy independence, $A$-dependence and the power law
were discussed.
A complementary confirmation of $z$-scaling for $\pi^0$-meson
and charged hadron production in $pp$ collisions at RHIC was obtained.

New measurements of the multiplicity charged particle density
and the inclusive cross sections of particle production
in the  experimentally inaccessible
kinematic region for the study of $z$-scaling were suggested.
The change of the anomalous fractal dimension ("$\delta$-jump") was suggested
to consider as a new and complementary signature of new physics phenomena
of high-$p_T$ particle production in hadron-hadron and hadron-nucleus
collisions at high energies.
The  $z-p_T$ plot was used to determine the regions that are
of more preferable for experimental search for $z$-scaling violation.
The properties of data $z$-presentation were used to predict
high-$p_T$ particle  spectra at RHIC and LHC energies.

\vskip 5mm
{\large \bf Acknowledgments}

 The author would like to thank
I.Zborovsk\'{y}, Yu.Panebratsev, O.Rogachevski and D.Toivonen
for collaboration and useful and stimulating  discussions of the problem.

\vskip 1cm

{\small

\newpage
\begin{minipage}{4cm}

\end{minipage}

\vskip 2cm
\begin{center}
\hspace*{-1.5cm}
\parbox{3.5cm}{\epsfxsize=3.5cm\epsfysize=3.5cm\epsfbox[95 95 400 400]
{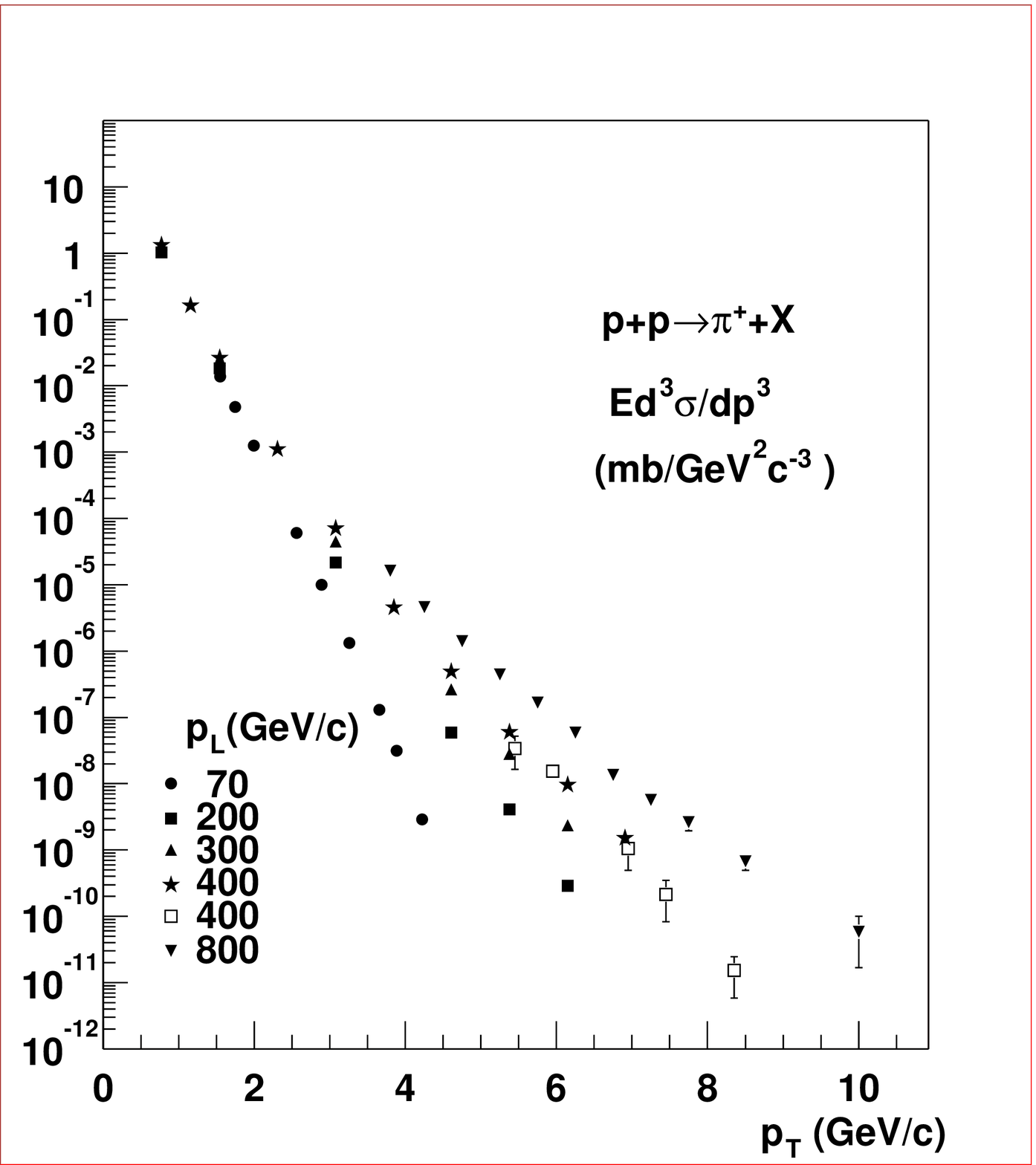}{}}
\hspace*{4cm}
\parbox{3.5cm}{\epsfxsize=3.5cm\epsfysize=3.5cm\epsfbox[95 95 400 400]
{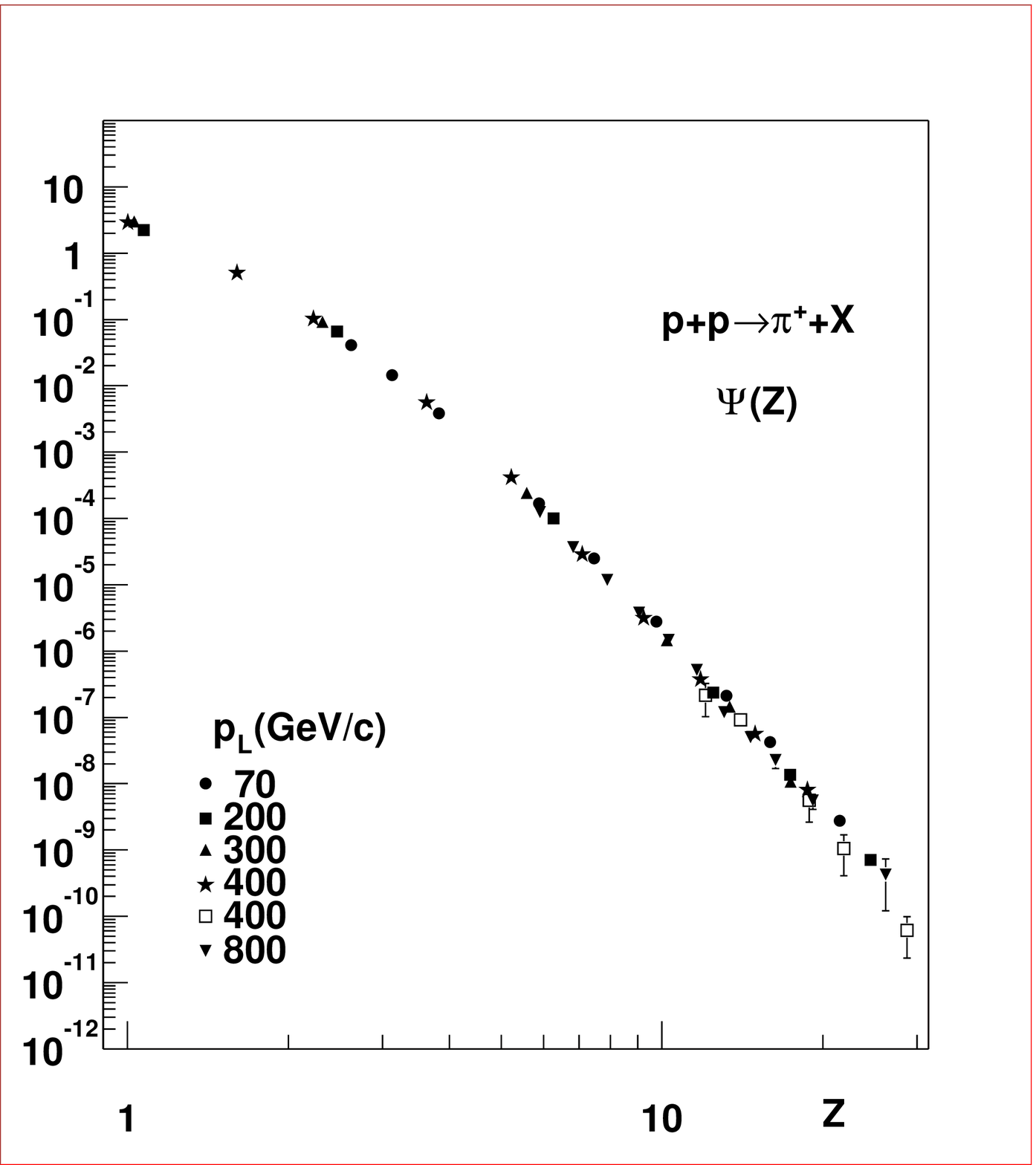}{}}
\vskip 1.5cm
\hspace*{0.cm} a) \hspace*{8.cm} b)\\[0.5cm]
\end{center}

{\bf Figure 1.} (a) The inclusive differential cross sections for  $\pi^+$-mesons produced in $pp$
collisions at $p_{lab} = 70, 200, 300,400$ and  800~GeV/c and $\theta_{cm} \simeq 90^{0}$ as functions of
the transverse momentum  $p_{T}$.
 Experimental data are taken from
\cite{Protvino,Cronin,Jaffe}. (b) The corresponding scaling function $\psi(z)$.
\vskip 4cm

\begin{center}
\hspace*{-1.5cm}
\parbox{3.5cm}{\epsfxsize=3.5cm\epsfysize=3.5cm\epsfbox[95 95 400 400]
{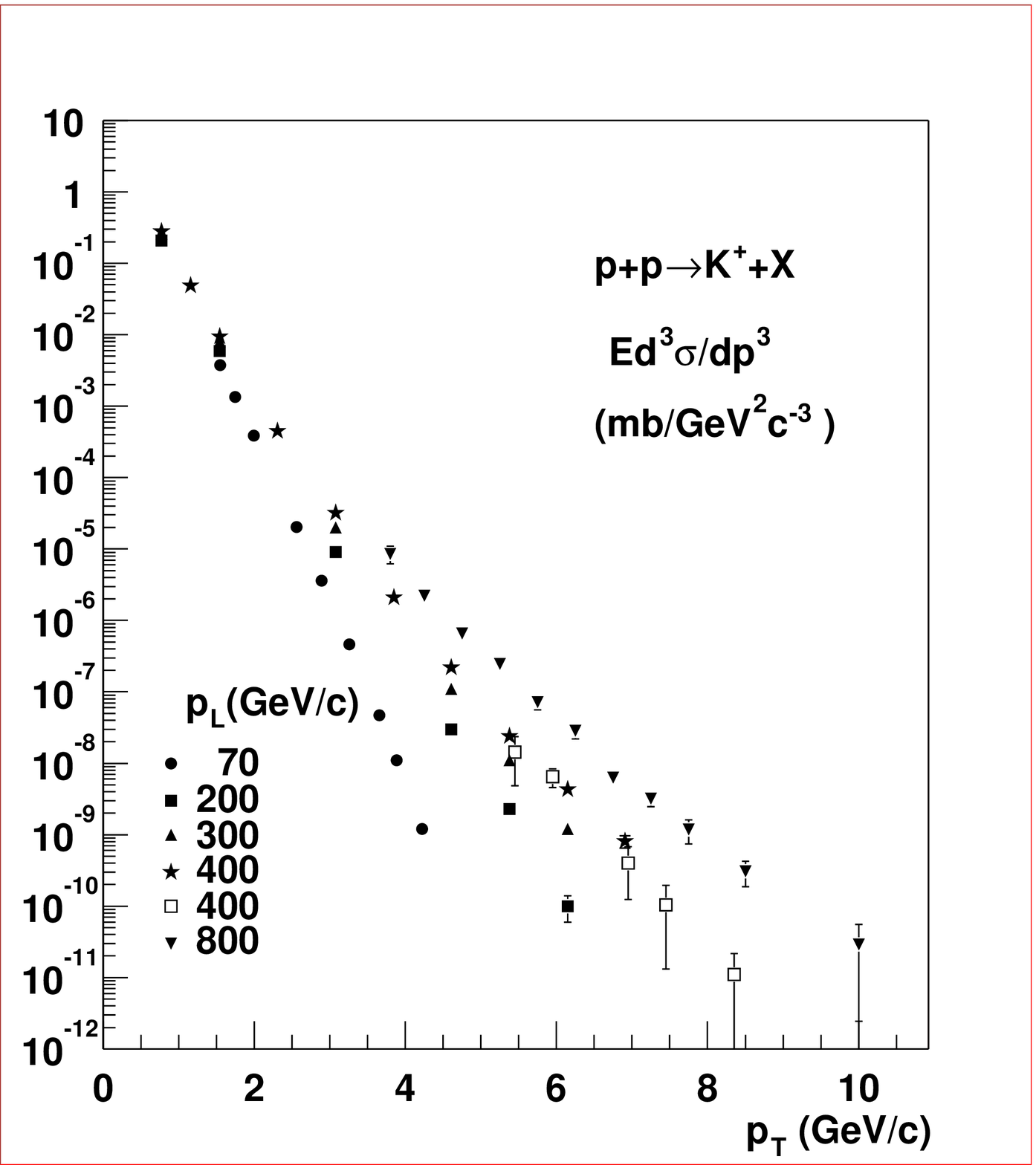}{}}
\hspace*{4cm}
\parbox{3.5cm}{\epsfxsize=3.5cm\epsfysize=3.5cm\epsfbox[95 95 400 400]
{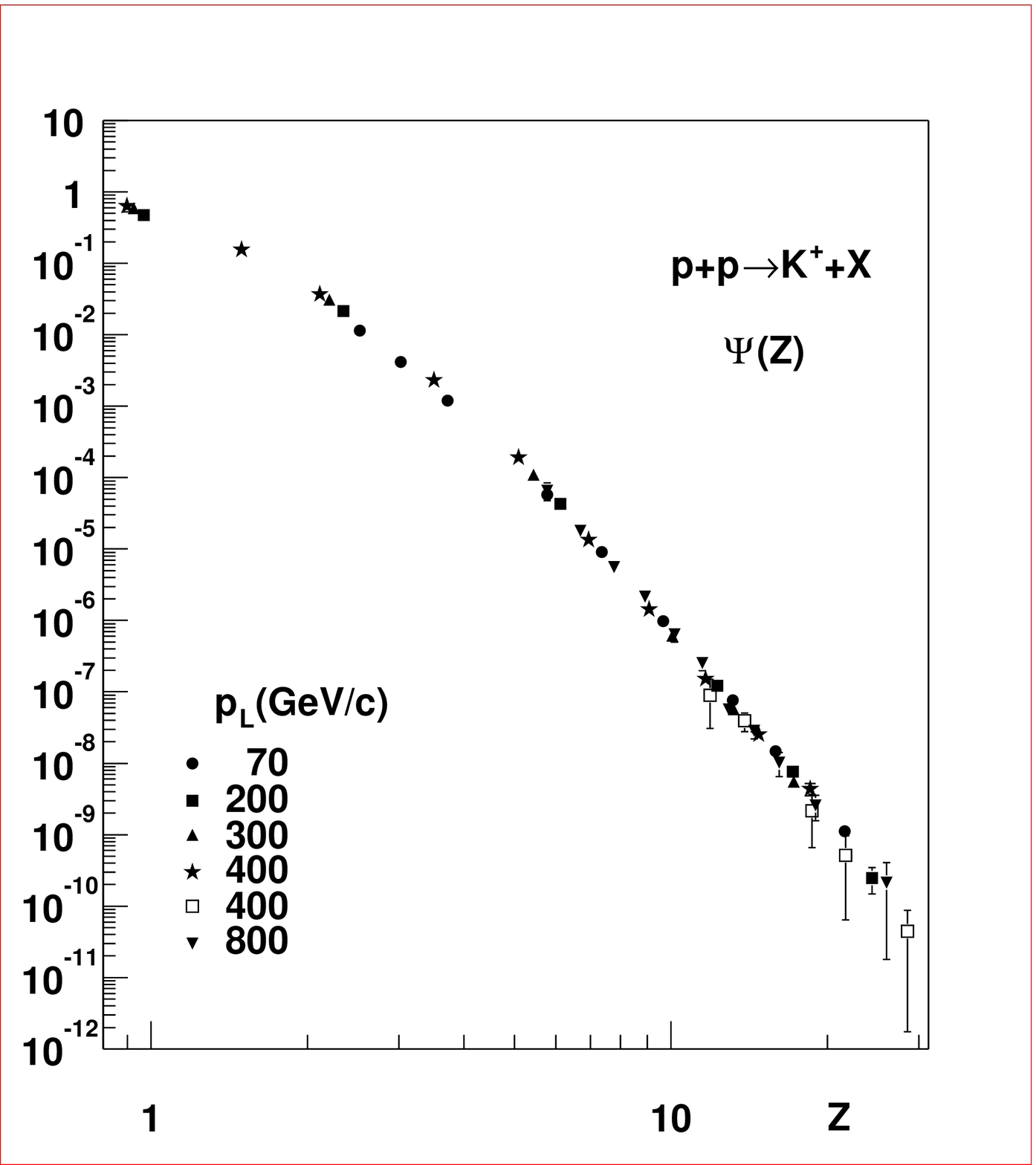}{}}
\vskip 1.5cm
\hspace*{0.cm} a) \hspace*{8.cm} b)\\[0.5cm]
\end{center}

{\bf Figure 2.} (a) The inclusive differential cross sections for
 $K^+$-mesons produced in $pp$ collisions
at $p_{lab} = 70, 200, 300,400$ and  800~GeV/c and $\theta_{cm} = 90^{0}$
as functions of the transverse momentum  $p_{T}$.
 Experimental data are taken from \cite{Protvino,Cronin,Jaffe}.
 (b) The corresponding scaling function $\psi(z)$.

\newpage
\begin{minipage}{4cm}

\end{minipage}

\vskip 4cm
\begin{center}
\hspace*{-2.5cm}
\parbox{5cm}{\epsfxsize=5.cm\epsfysize=5.cm\epsfbox[95 95 400 400]
{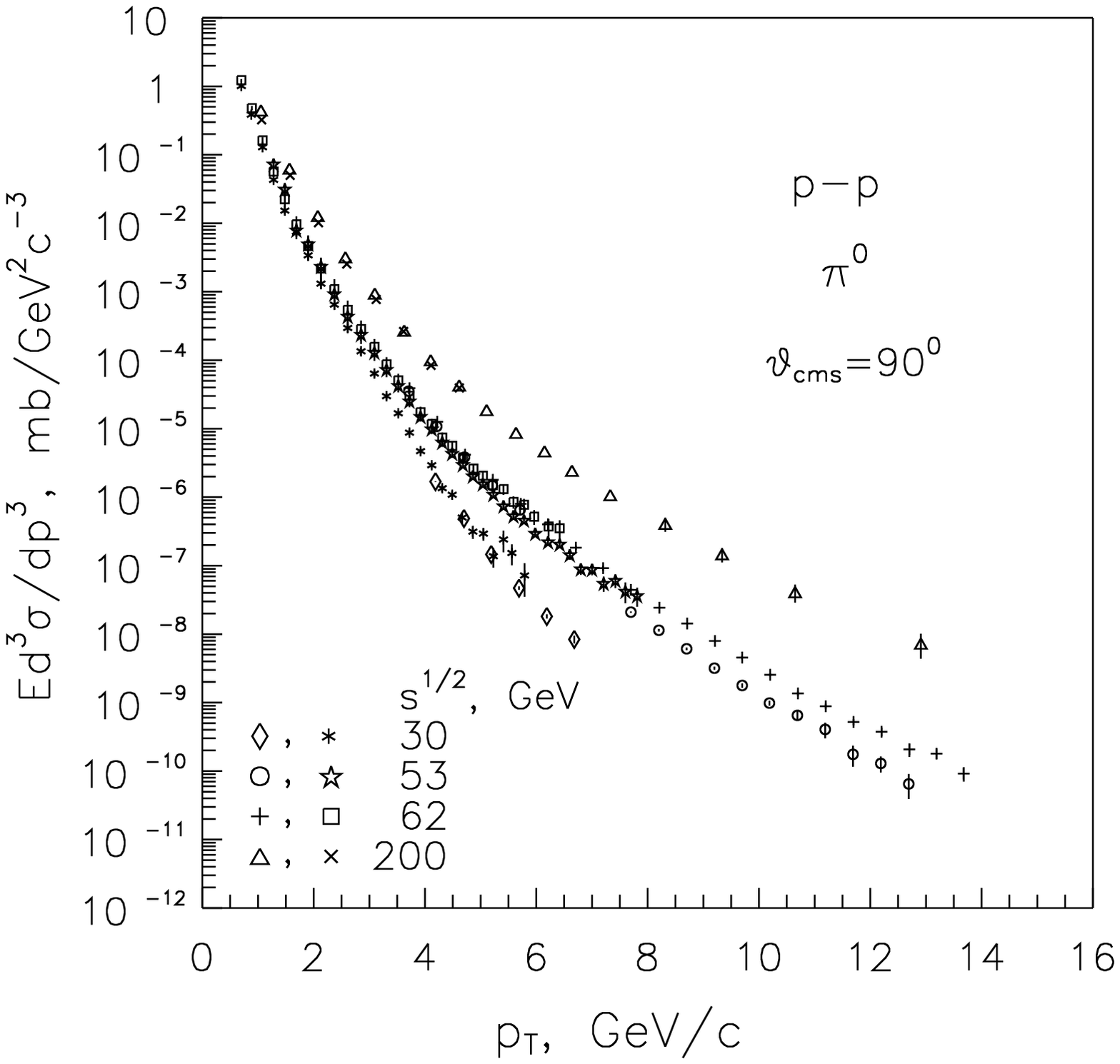}{}}
\hspace*{3cm}
\parbox{5cm}{\epsfxsize=5.cm\epsfysize=5.cm\epsfbox[95 95 400 400]
{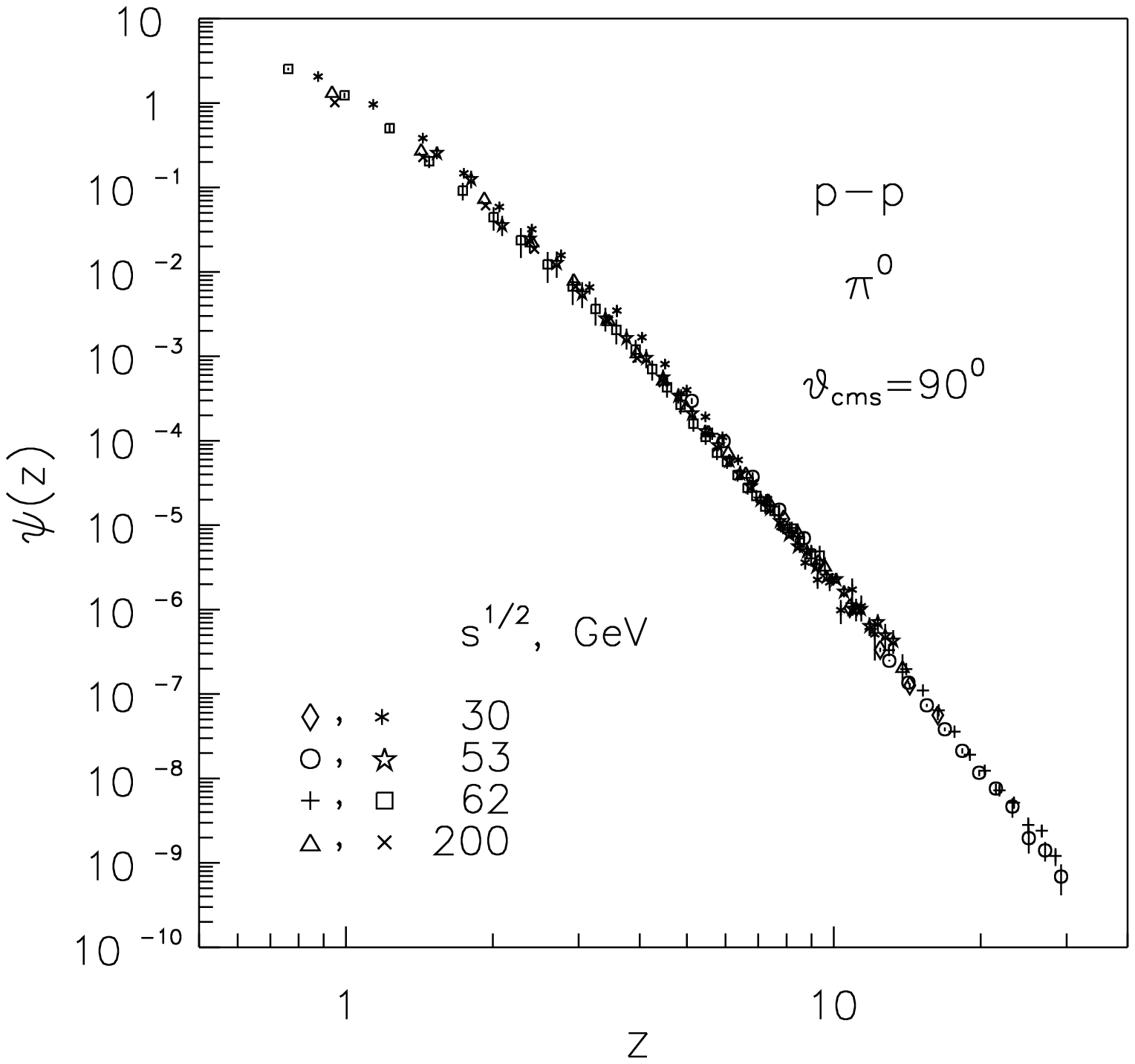}{}}
 \vskip -0.5cm
\hspace*{0.cm} a) \hspace*{8.cm} b)\\[0.5cm]
\end{center}

{\bf Figure 3.}
(a) The dependence of  the inclusive cross section of $\pi^0$-meson production
on the transverse
momentum $p_{\bot}$ in $pp$  collisions at $\sqrt s = 30,53,62$ and 200~GeV
and an angle $\theta_{cm}$ of $90^0$.
The experimental data  are taken from
\cite{Phenix}-\cite{Eggert}.
(b) The corresponding scaling function $\psi(z)$.

\vskip 5cm

\begin{center}
\hspace*{-2.5cm}
\parbox{5cm}{\epsfxsize=5.cm\epsfysize=5.cm\epsfbox[95 95 400 400]
{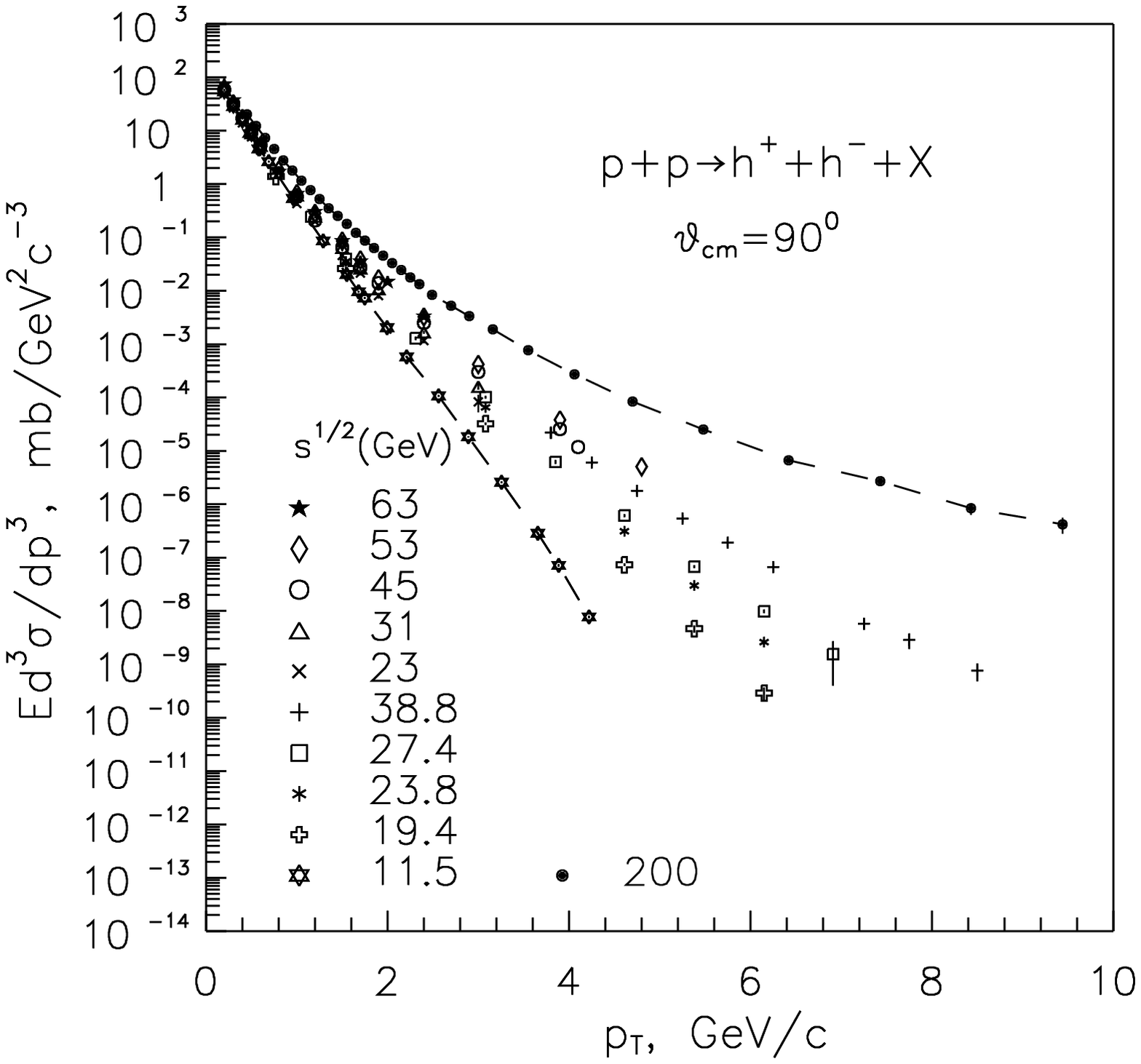}{}}
\hspace*{3cm}
\parbox{5cm}{\epsfxsize=5.cm\epsfysize=5.cm\epsfbox[95 95 400 400]
{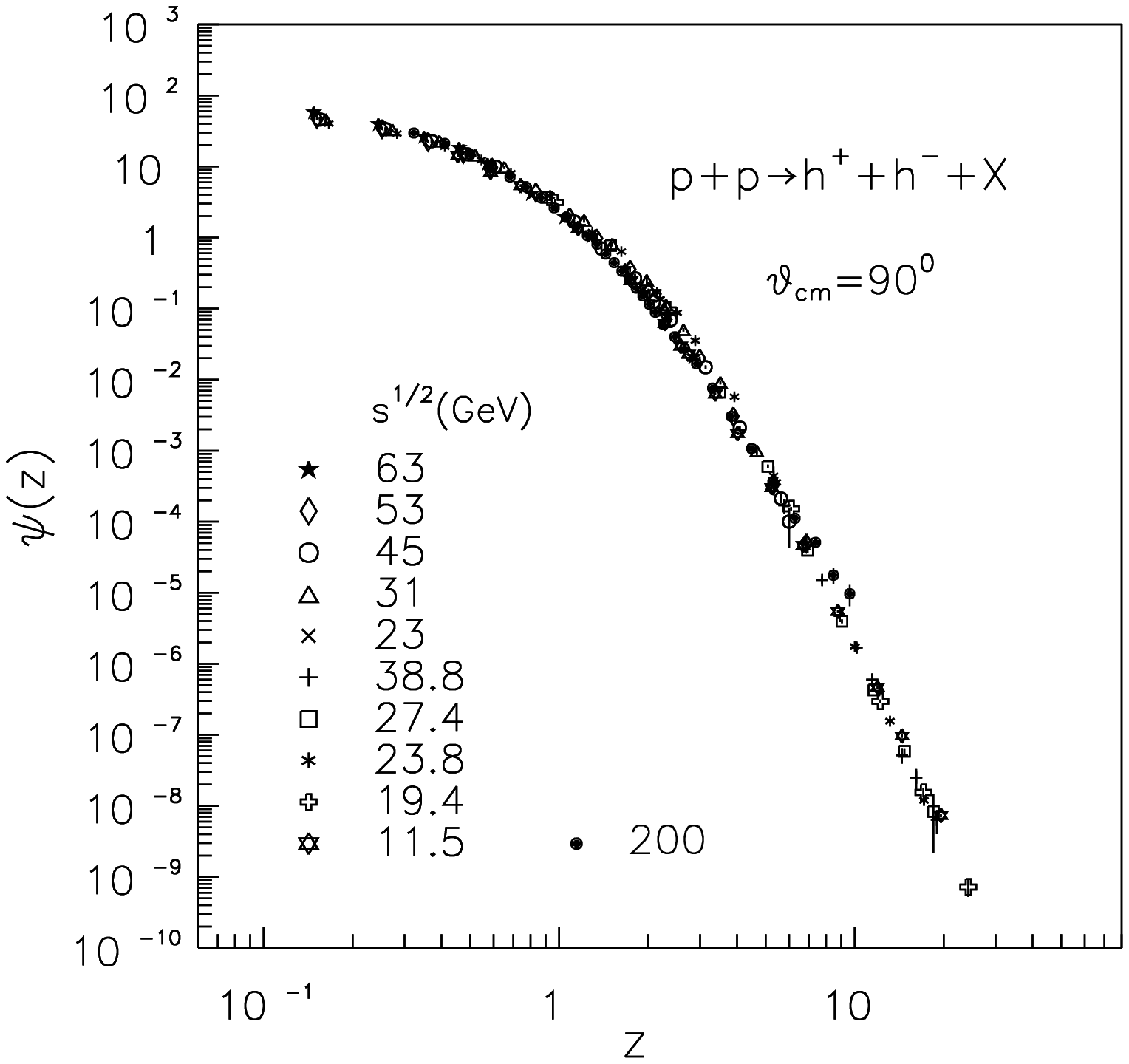}{}}
\vskip -1.cm
\hspace*{0.cm} a) \hspace*{8.cm} b)\\[0.5cm]
\end{center}

{\bf Figure 4.}
 (a) Inclusive cross sections of charged hadron production in $pp$ collisions
 versus transverse momentum  at U70, ISR, Tevatron  and  RHIC energies
 $\sqrt s = (11.5-200)$ GeV  and an angle $\theta_{cm} $ of $ 90^{0}$.
 Experimental data are taken from
 \cite{Protvino,Cronin,Jaffe,STAR,Alper}.
 (b) The corresponding scaling function $\psi(z)$.}

\newpage
\begin{minipage}{4cm}

\end{minipage}

\vskip 4cm
\begin{center}
\hspace*{-2.5cm}
\parbox{5cm}{\epsfxsize=5.cm\epsfysize=5.cm\epsfbox[95 95 400 400]
{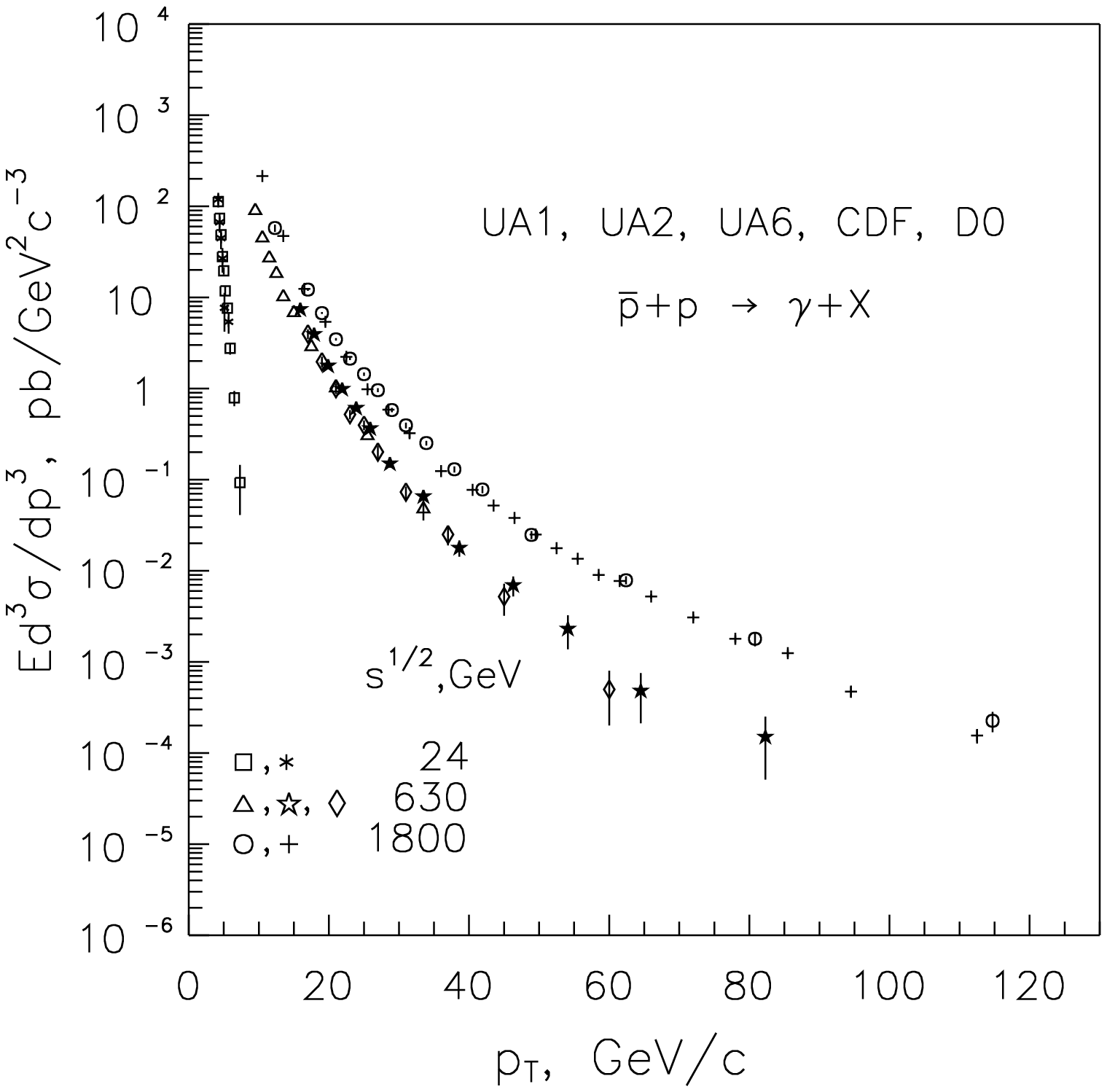}{}}
\hspace*{3cm}
\parbox{5cm}{\epsfxsize=5.cm\epsfysize=5.cm\epsfbox[95 95 400 400]
{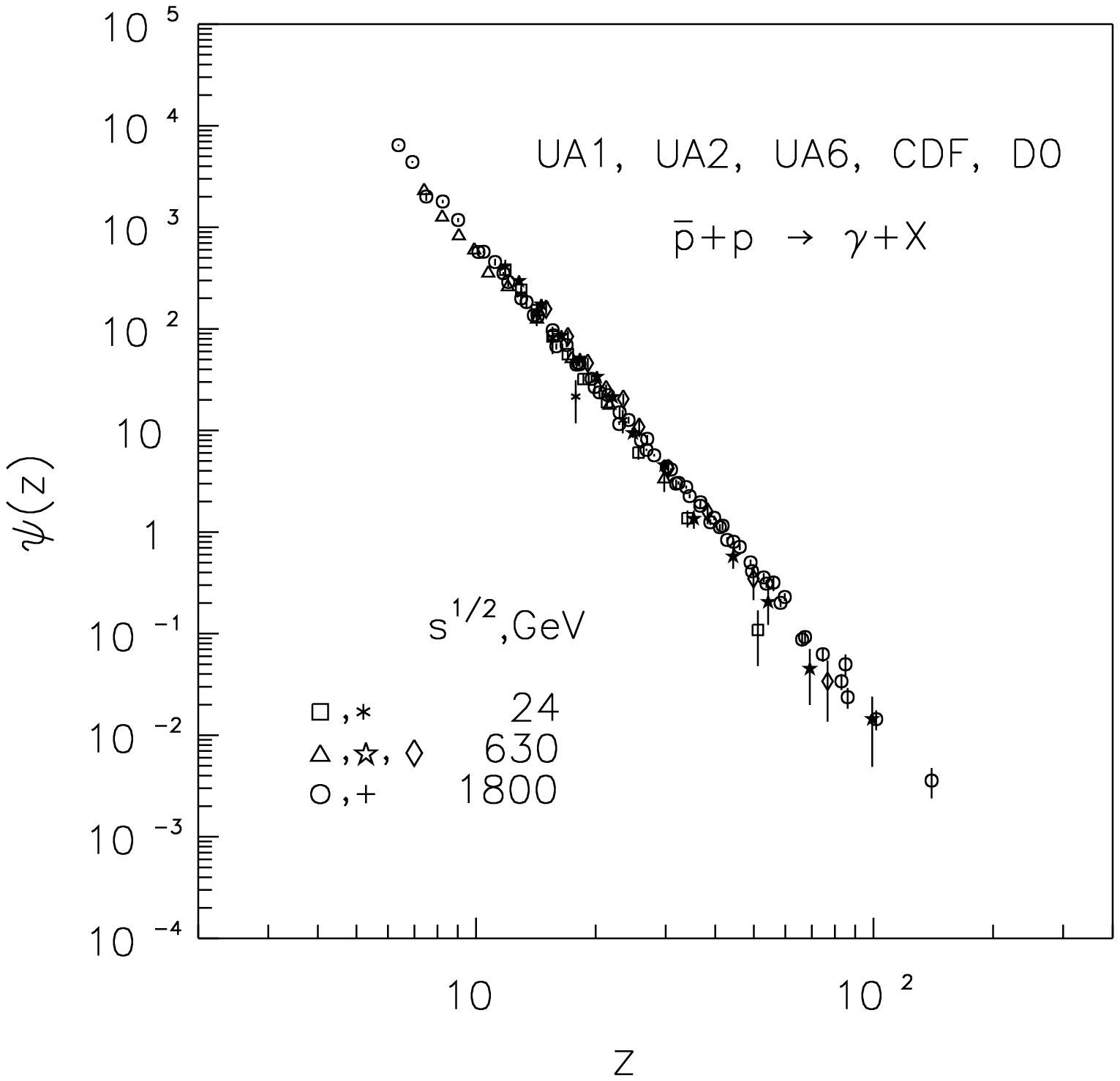}{}}
\vskip -0.5cm
\hspace*{0.cm} a) \hspace*{8.cm} b)\\[0.5cm]
\end{center}

{\bf Figure 5.}
(a) The dependence of the inclusive cross section of direct photon
production on the transverse momentum $p_T$ in $\bar pp$
collisions at $\sqrt s = (24-1800)$ GeV. Experimental data obtained by the UA1,
UA2, UA6, CDF and D0 Collaborations are
taken from
\cite{UA1}-\cite{D0pho}.
(b) The corresponding scaling functions.

\vskip 5cm

\begin{center}
\hspace*{-2.5cm}
\parbox{5cm}{\epsfxsize=5.cm\epsfysize=5.cm\epsfbox[95 95 400 400]
{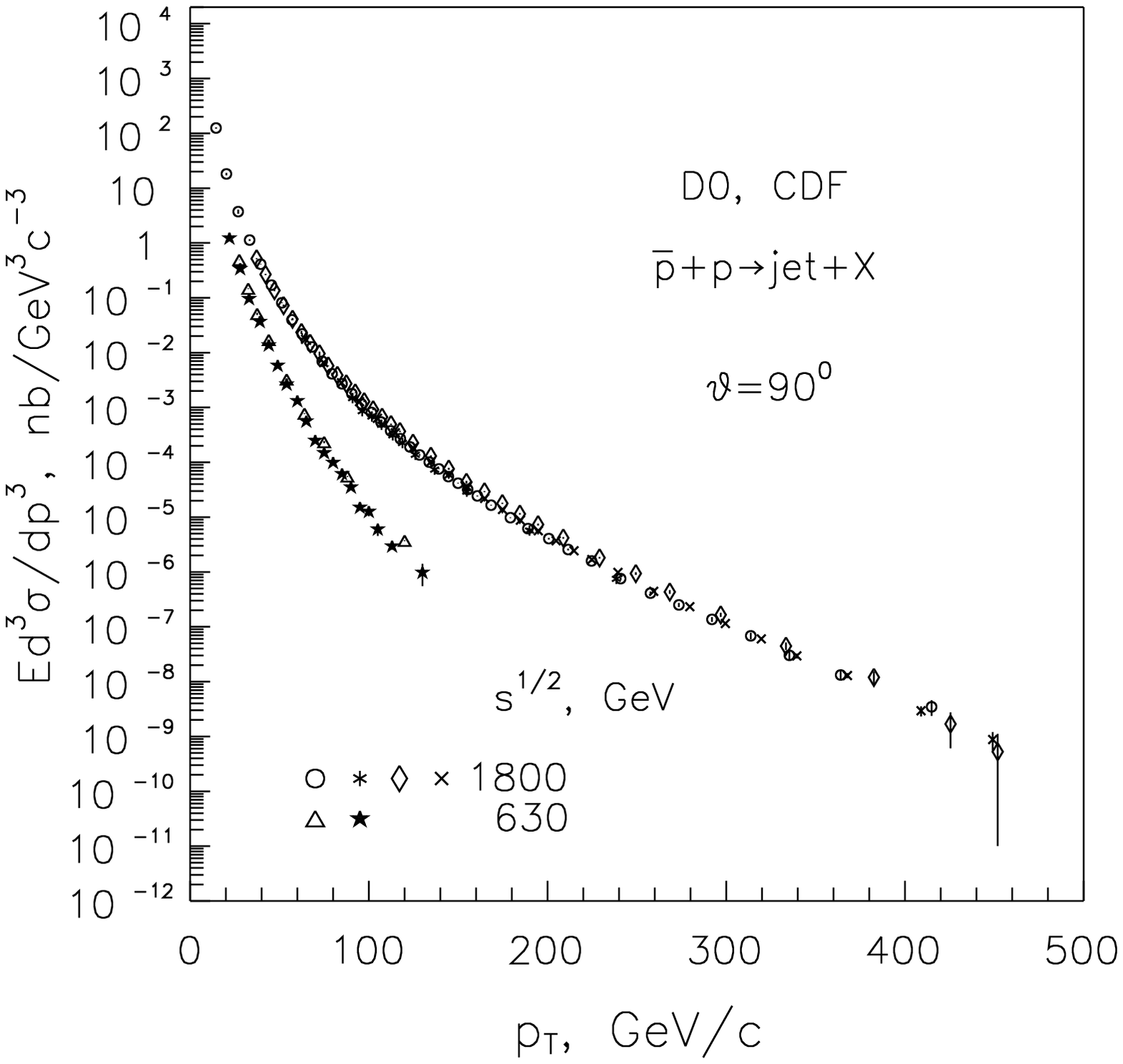}{}}
\hspace*{3cm}
\parbox{5cm}{\epsfxsize=5.cm\epsfysize=5.cm\epsfbox[95 95 400 400]
{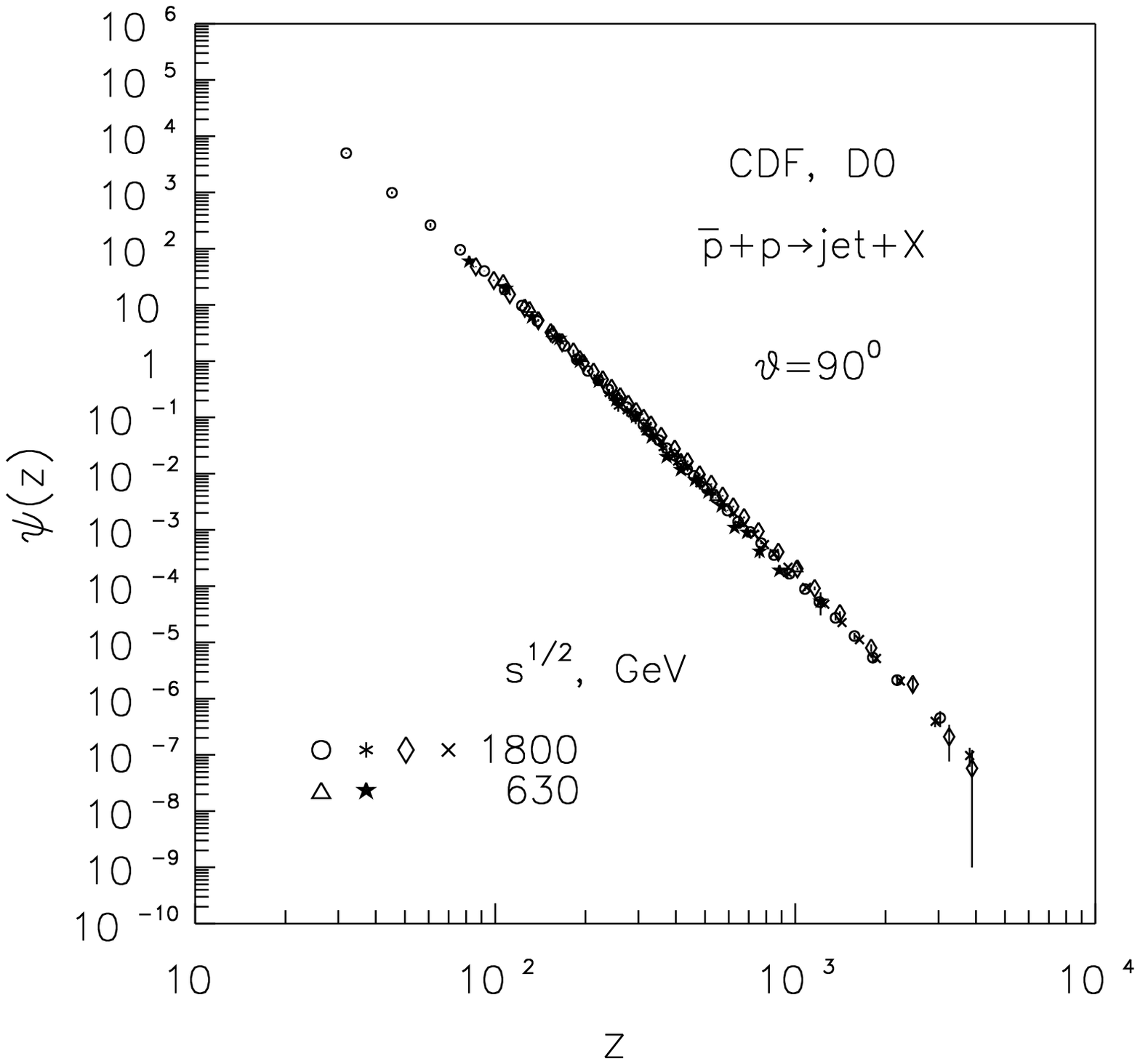}{}}
\vskip -1.cm
\hspace*{0.cm} a) \hspace*{8.cm} b)\\[0.5cm]
\end{center}

{\bf Figure 6.}
(a) Inclusive cross sections of jet production in $\bar pp$ collisions
 versus transverse momentum  at Tevatron energies $\sqrt s = 630$ and
 $1800$~GeV and $\theta_{cm} \simeq 90^{0}$ obtained by CDF and D0
 Collaborations. Experimental data are taken from  \cite{CDFjet,D0jet}.
 (b) The corresponding scaling function $\psi(z)$.

\newpage
\begin{minipage}{4cm}

\end{minipage}

\vskip 4cm
\begin{center}
\hspace*{-2.5cm}
\parbox{5cm}{\epsfxsize=5.cm\epsfysize=5.cm\epsfbox[95 95 400 400]
{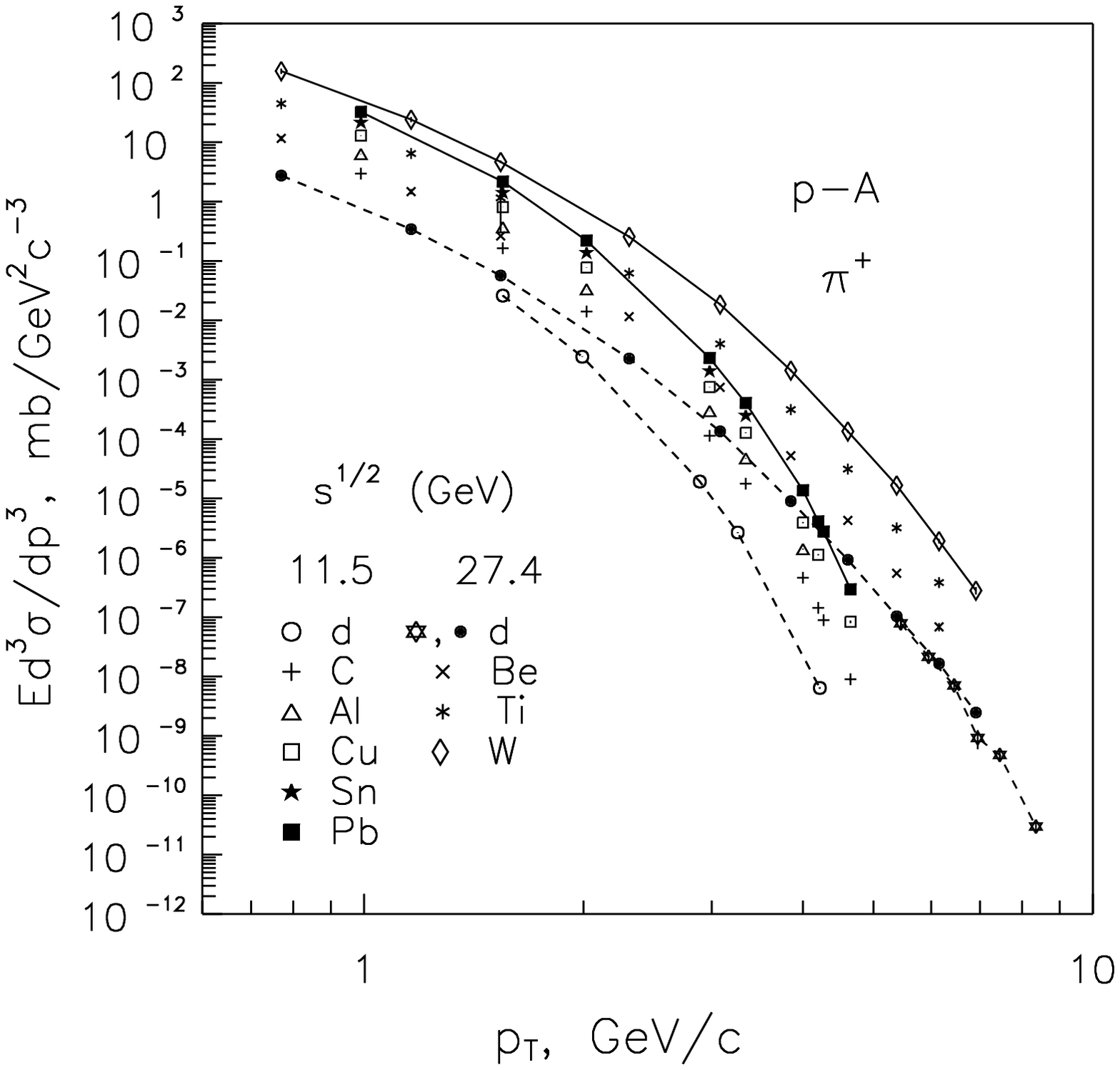}{}}
\hspace*{3cm}
\parbox{5cm}{\epsfxsize=5.cm\epsfysize=5.cm\epsfbox[95 95 400 400]
{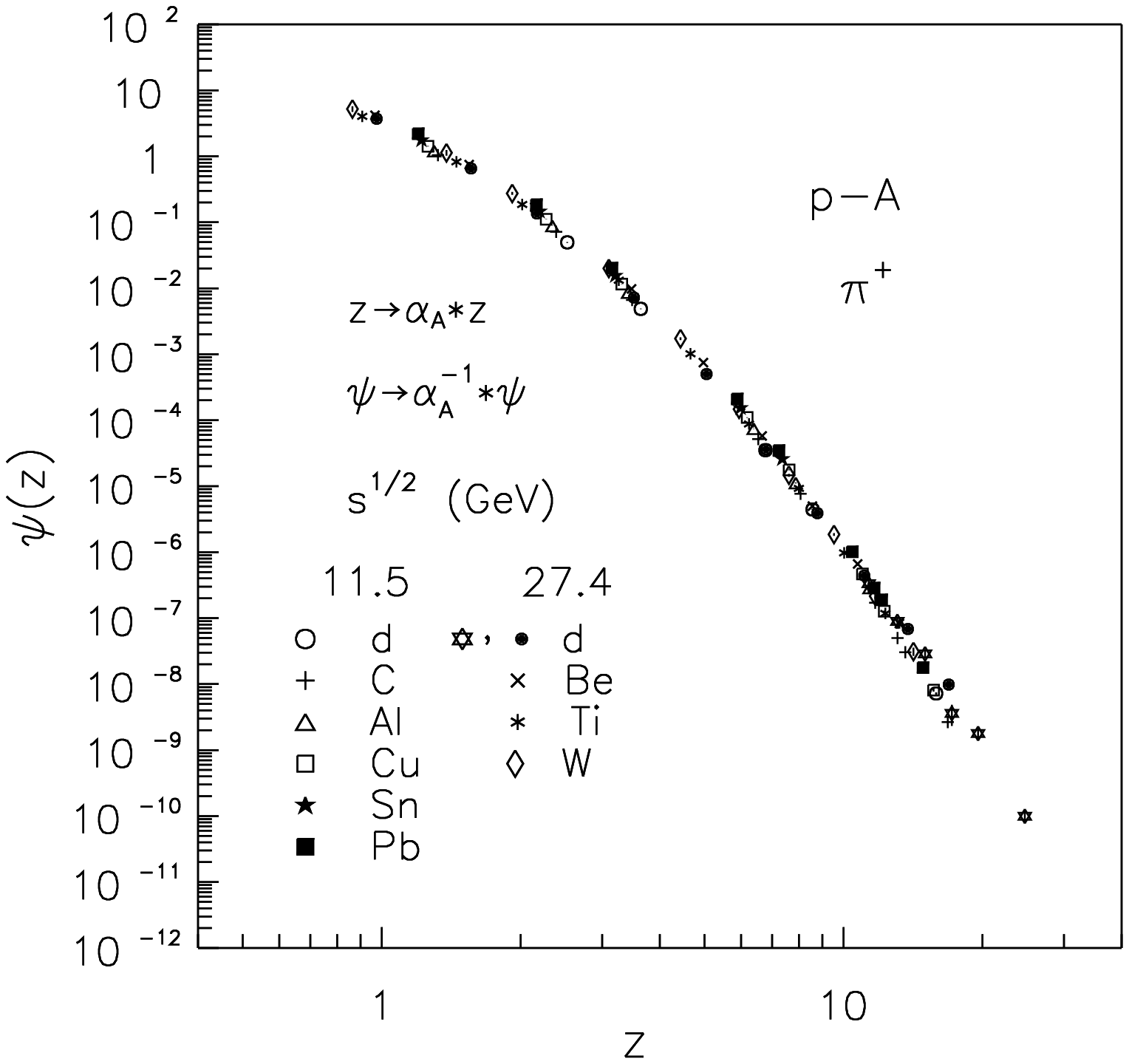}{}}
\vskip -0.5cm
\hspace*{0.cm} a) \hspace*{8.cm} b)\\[0.5cm]
\end{center}

{\bf Figure 7.}
 (a) Inclusive differential cross section
 for  $\pi^+$-mesons produced
in $pA$ collisions  at $\sqrt s  = 11.5$ and 27.4~GeV/c
and  $\theta_{cm}^{NN} \simeq 90^{0}$
as a function of the transverse momentum  $p_T$.
Solid and dashed lines are obtained by fitting  of the data
for $W, Pb$ and $D$, respectively. Experimental data are taken from
\cite{Protvino,Cronin,Jaffe}.
(b) The corresponding  scaling function $\psi(z)$.

\vskip 5cm

\begin{center}
\hspace*{-2.5cm}
\parbox{5cm}{\epsfxsize=5.cm\epsfysize=5.cm\epsfbox[95 95 400 400]
{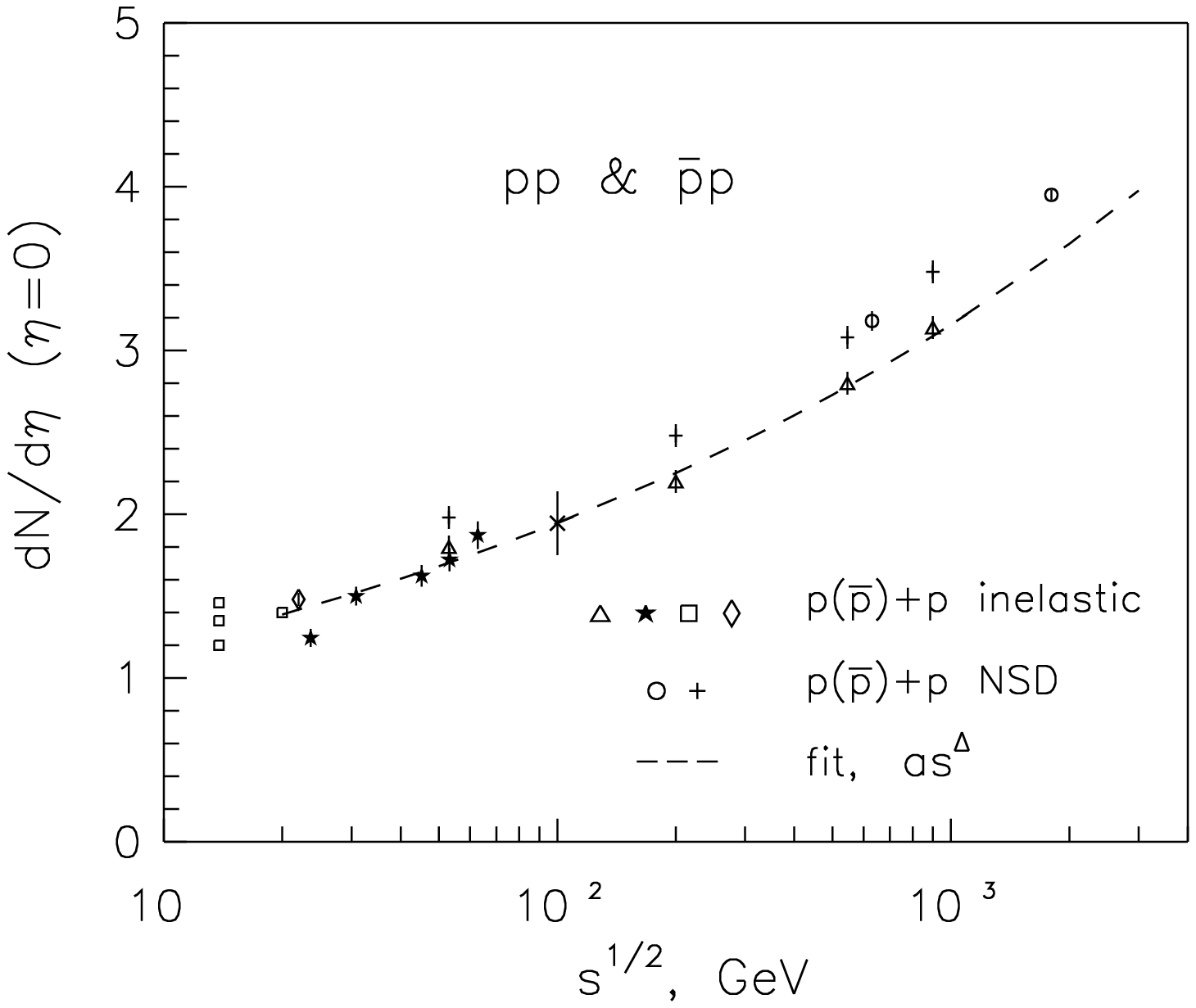}{}}
\hspace*{3cm}
\parbox{5cm}{\epsfxsize=5.cm\epsfysize=5.cm\epsfbox[95 95 400 400]
{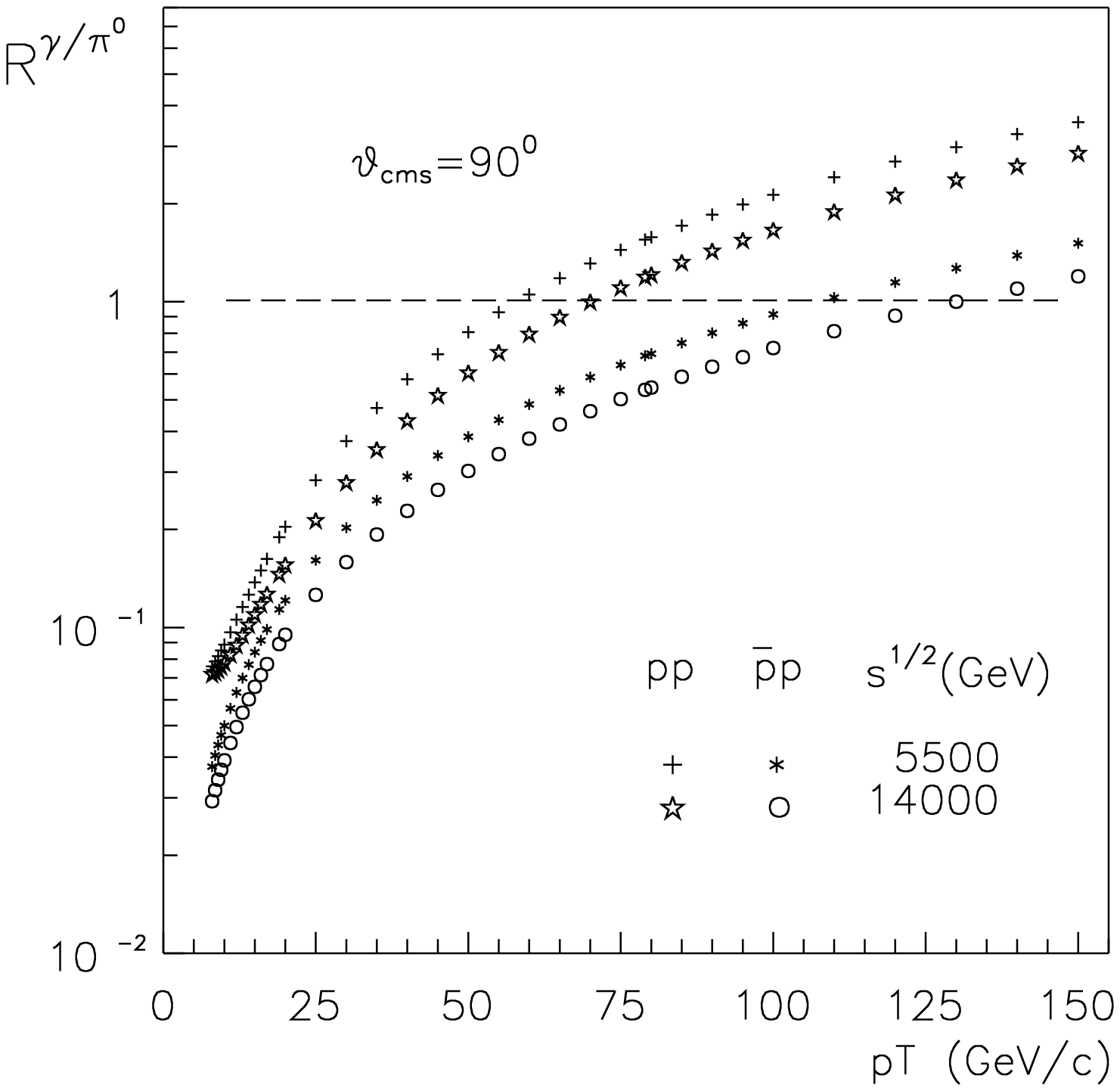}{}}
\vskip -1.cm
\hspace*{0.cm} a) \hspace*{8.cm} b)\\[0.5cm]
\end{center}

{\bf Figure 8.}
(a) The multiplicity charged particle density $dN/d\eta$ as a
function  of collision energy $\sqrt s $ at $\eta =0$ for $pp$ and
$\bar pp$ collisions. Experimental data are taken from
\cite{Thome}.
 (b) The $\gamma / \pi^0 $ ratio of inclusive cross sections
 versus the transverse momentum $p_T$  in $pp$ and $\bar pp$ collisions at
 $\sqrt s = 5.5$ and 14.~TeV.

\newpage
\begin{minipage}{4cm}

\end{minipage}

\vskip 2cm
\begin{center}
\hspace*{-1.5cm}
\parbox{3.5cm}{\epsfxsize=3.5cm\epsfysize=3.5cm\epsfbox[95 95 400 400]
{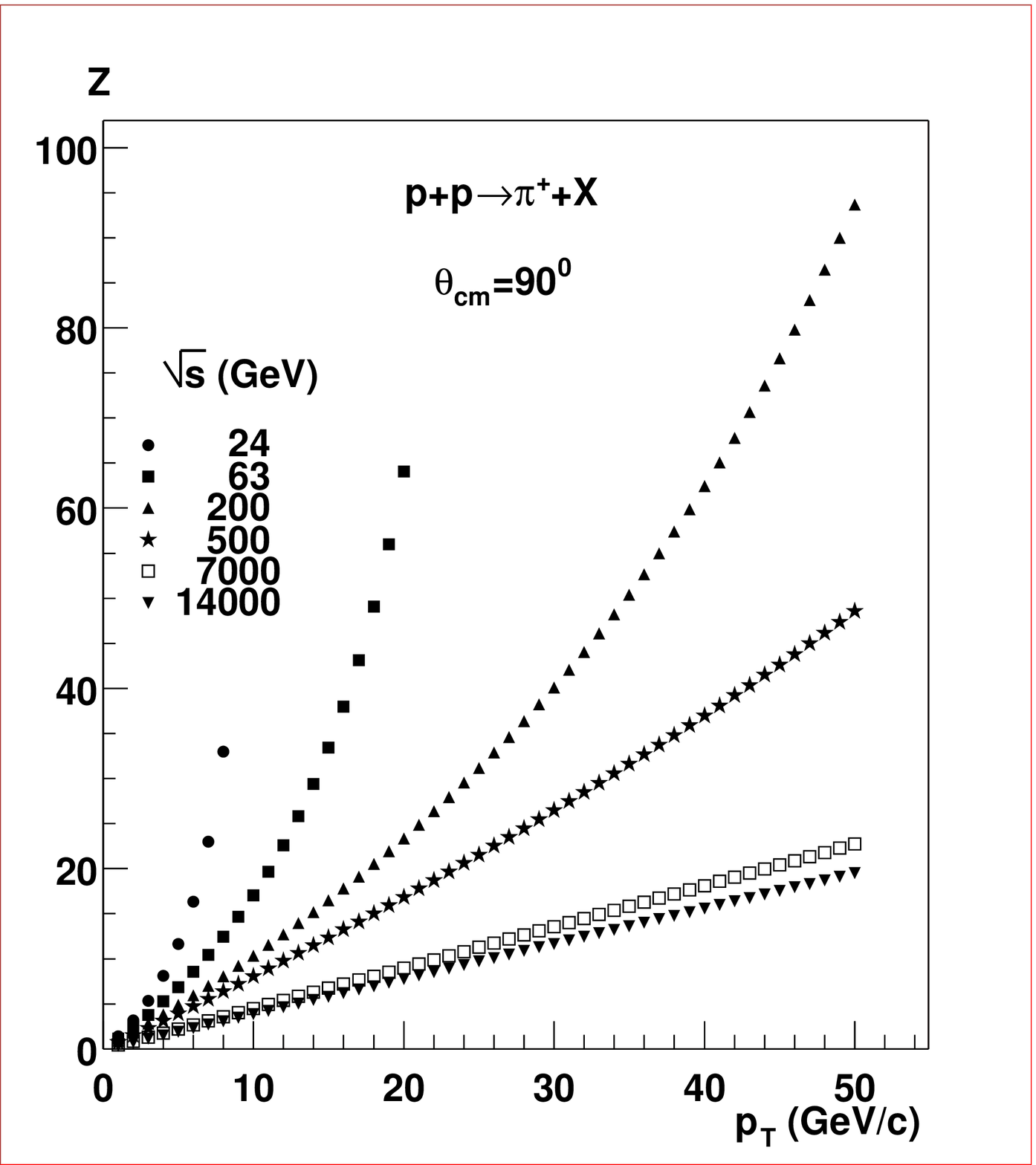}{}}
\hspace*{4cm}
\parbox{3.5cm}{\epsfxsize=3.5cm\epsfysize=3.5cm\epsfbox[95 95 400 400]
{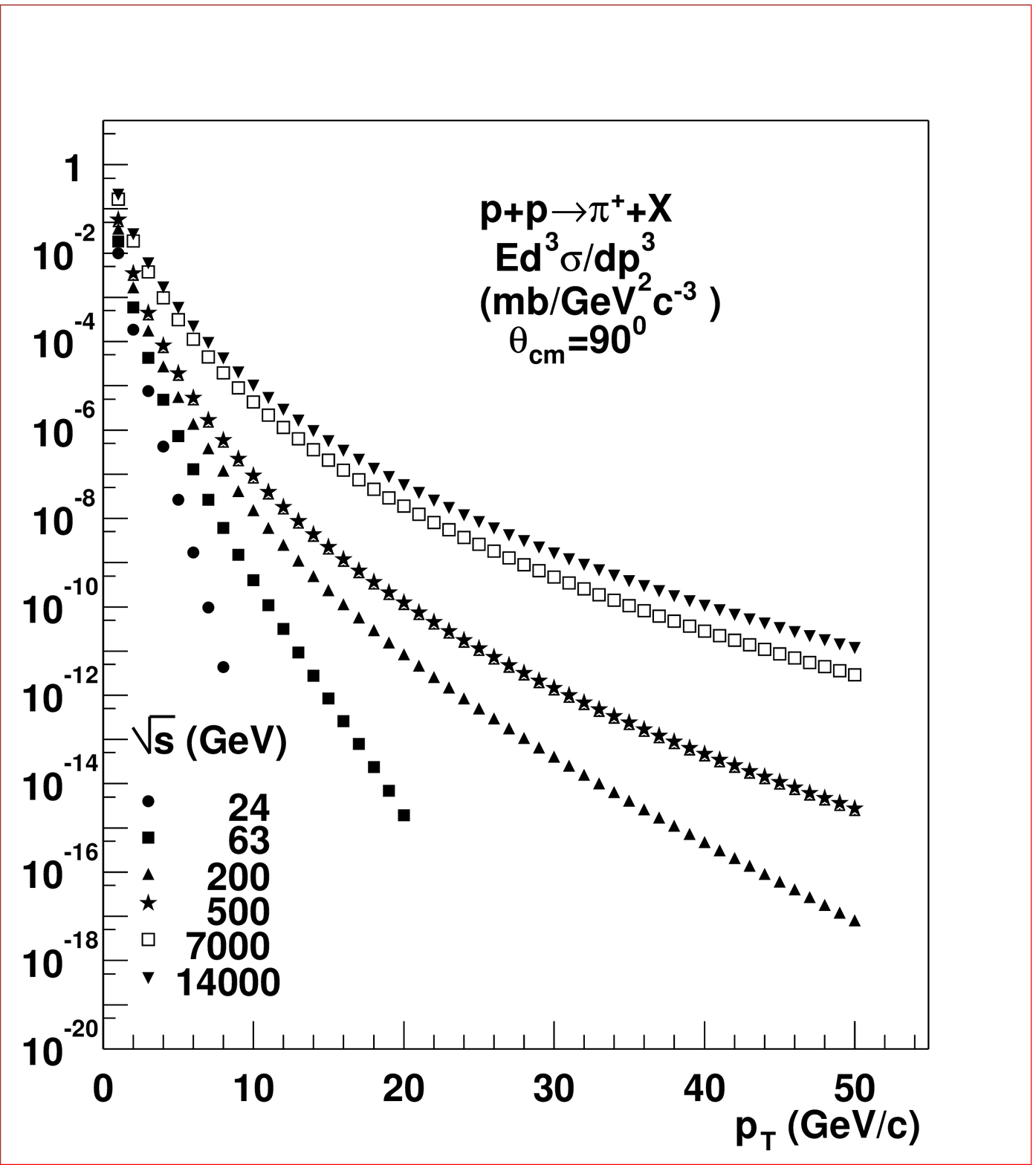}{}}
\vskip  1.5cm
\hspace*{0.cm} a) \hspace*{8.cm} b)\\[0.5cm]
\end{center}

{\bf Figure 9.} (a) The $z-p_T$ plot
and  (b) the dependence of  the inclusive cross section of
 $\pi^+$-meson production in $pp$ collisions
on the transverse momentum $p_T$
at $\sqrt s = (24-14000)$ GeV and an angle $\theta_{cm}$ of $90^0$.

\vskip 5cm

\begin{center}
\hspace*{-2.5cm}
\parbox{5cm}{\epsfxsize=5.cm\epsfysize=5.cm\epsfbox[95 95 400 400]
{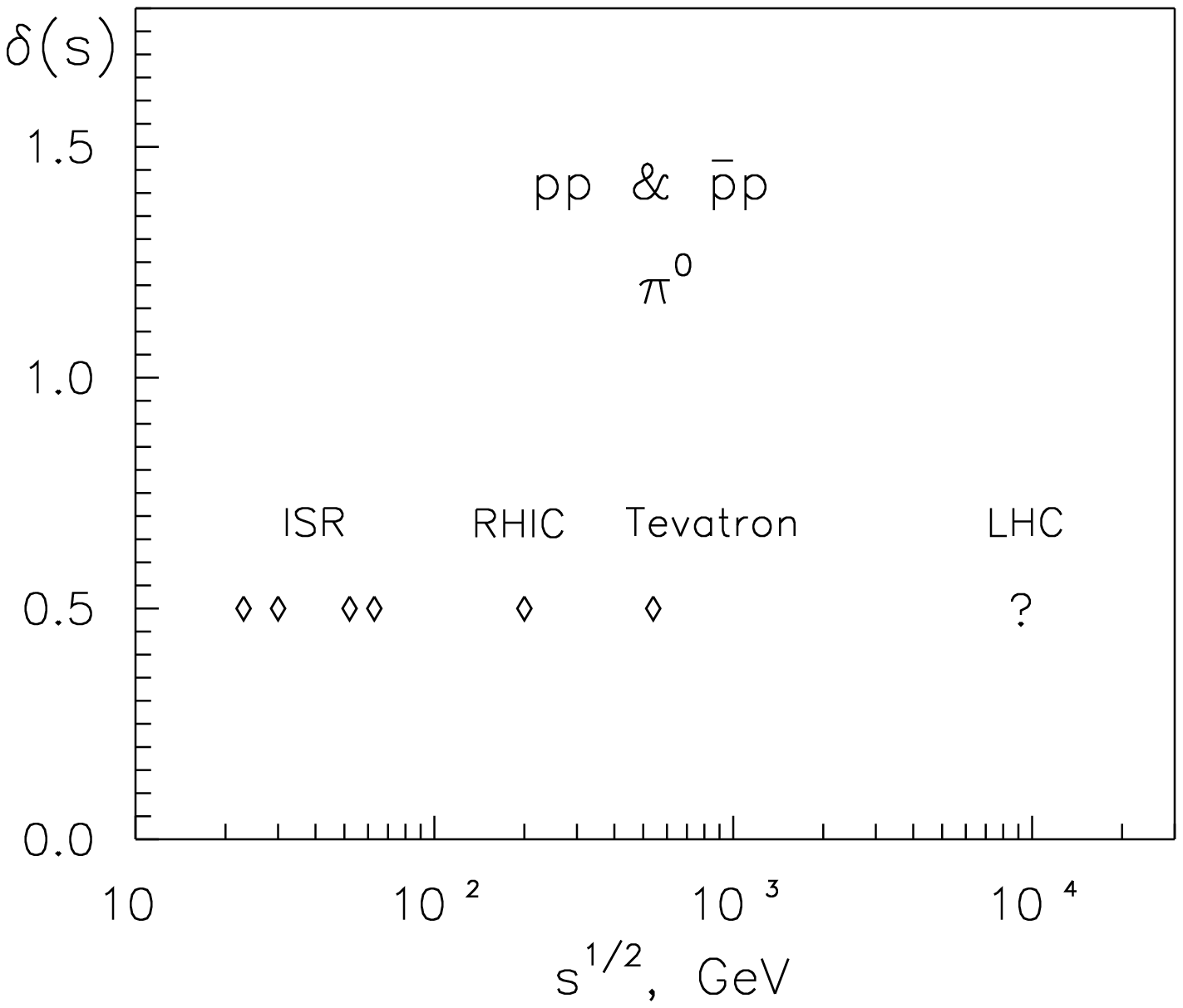}{}}
\hspace*{3cm}
\parbox{5cm}{\epsfxsize=5.cm\epsfysize=5.cm\epsfbox[95 95 400 400]
{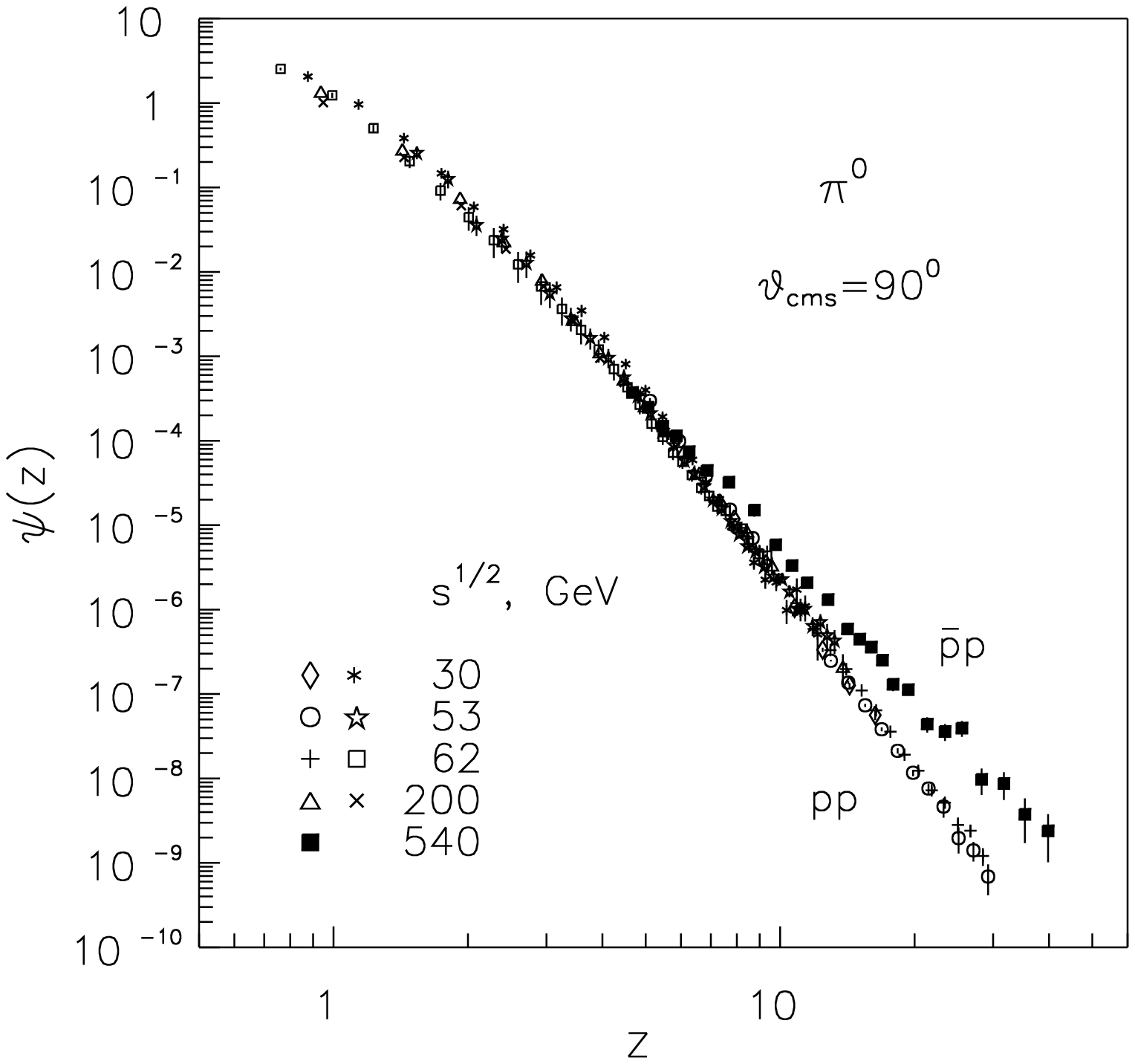}{}}
\vskip -1.cm
\hspace*{0.cm} a) \hspace*{8.cm} b)\\[0.5cm]
\end{center}

{\bf Figure 10.} (a) The dependence of  the anomalous fractal dimension
$\delta (s)$ on collision energy $\sqrt s $.
(b) The scaling function $\psi(z) $ of $\pi^0$-meson production  in  $pp$
and  $\bar pp$ collisions  on the transverse momentum $p_T$
at energy $\sqrt s = 30-200$ and 540~GeV and an angle $\theta_{cm}$
of $90^0$, respectively.
The experimental data  are taken from
\cite{Phenix}-\cite{Eggert}
 and \cite{Banner}.

\newpage
\begin{minipage}{4cm}

\end{minipage}

\vskip 4cm
\begin{center}
\hspace*{-2.5cm}
\parbox{5cm}{\epsfxsize=5.cm\epsfysize=5.cm\epsfbox[95 95 400 400]
{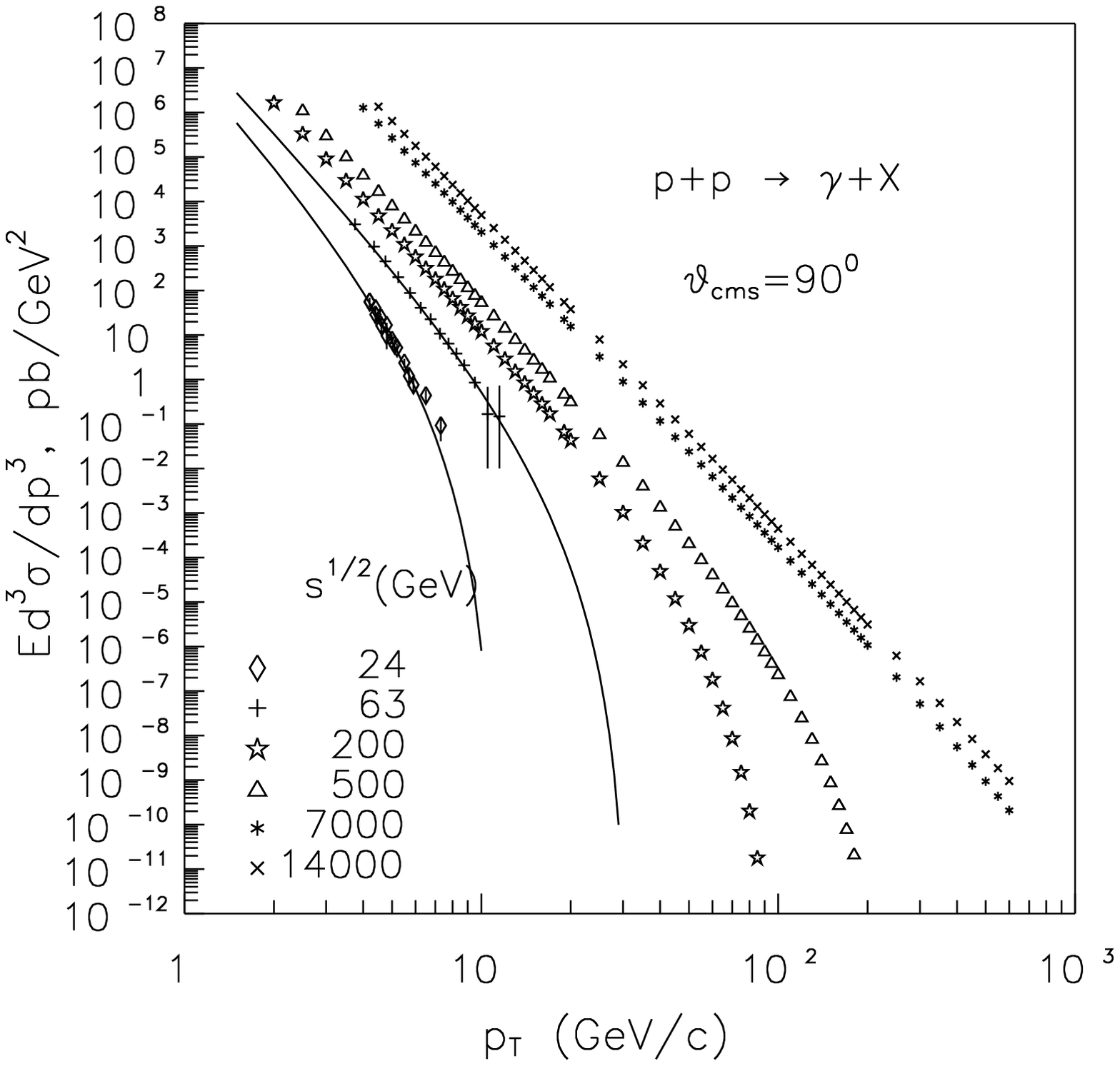}{}}
\hspace*{3cm}
\parbox{5cm}{\epsfxsize=5.cm\epsfysize=5.cm\epsfbox[95 95 400 400]
{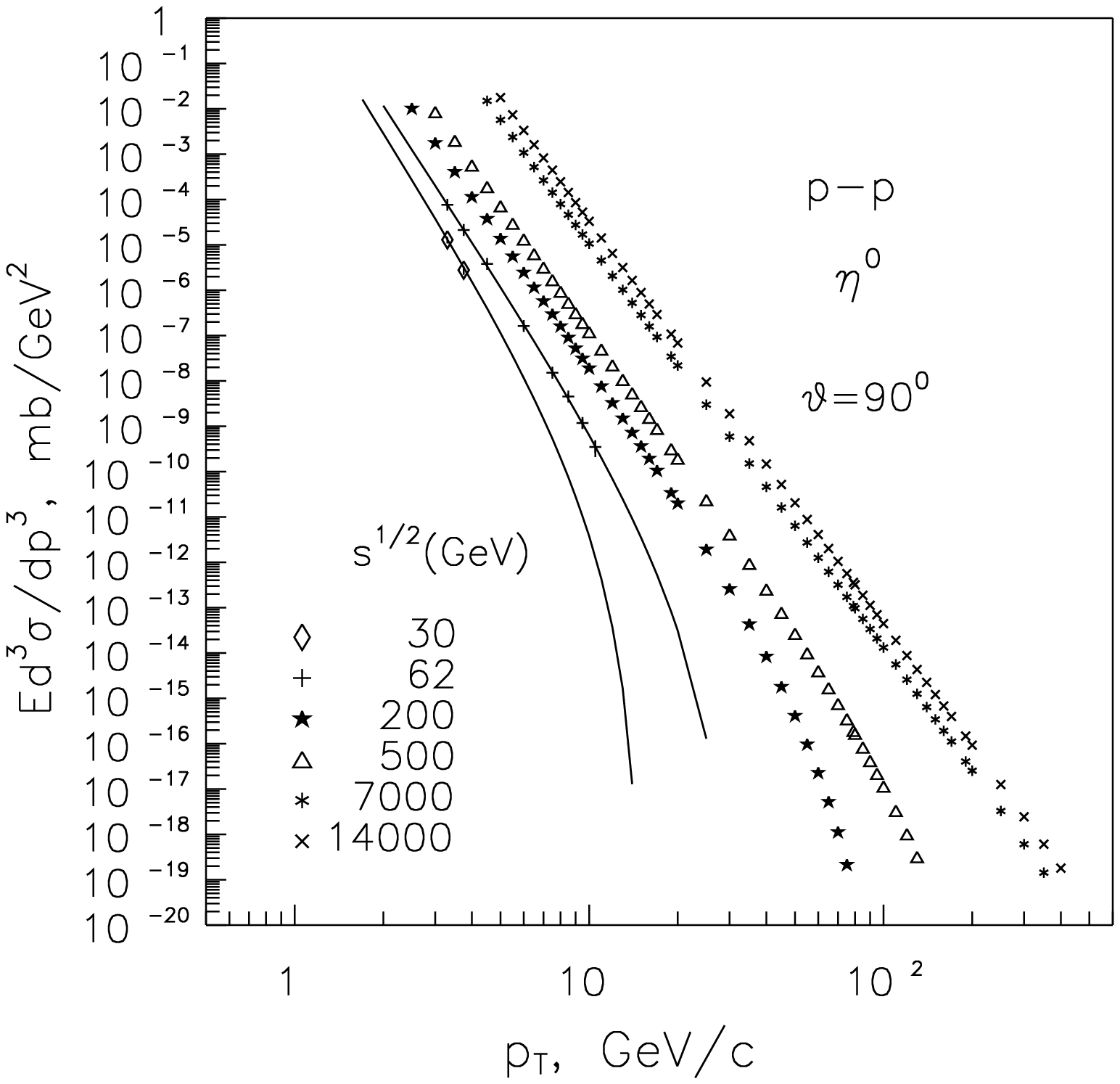}{}}
\vskip -0.5cm
\hspace*{0.cm} a) \hspace*{8.cm} b)\\[0.5cm]
\end{center}

{\bf Figure 11.}
 The dependence of inclusive cross sections of direct photon (a)
 and $\eta^0$-meson (b)  production on the
transverse momentum $p_T$ in $pp$ collisions at $\sqrt s =
(24-14000)$ GeV. Experimental data are taken from
\cite{Kou2,UA6p,R806}. Solid lines and points ($\star,
\triangle, *, \times$) are the calculated results.

\vskip 5cm

\begin{center}
\hspace*{-2.5cm}
\parbox{5cm}{\epsfxsize=5cm\epsfysize=5cm\epsfbox[95 95 400 400]
{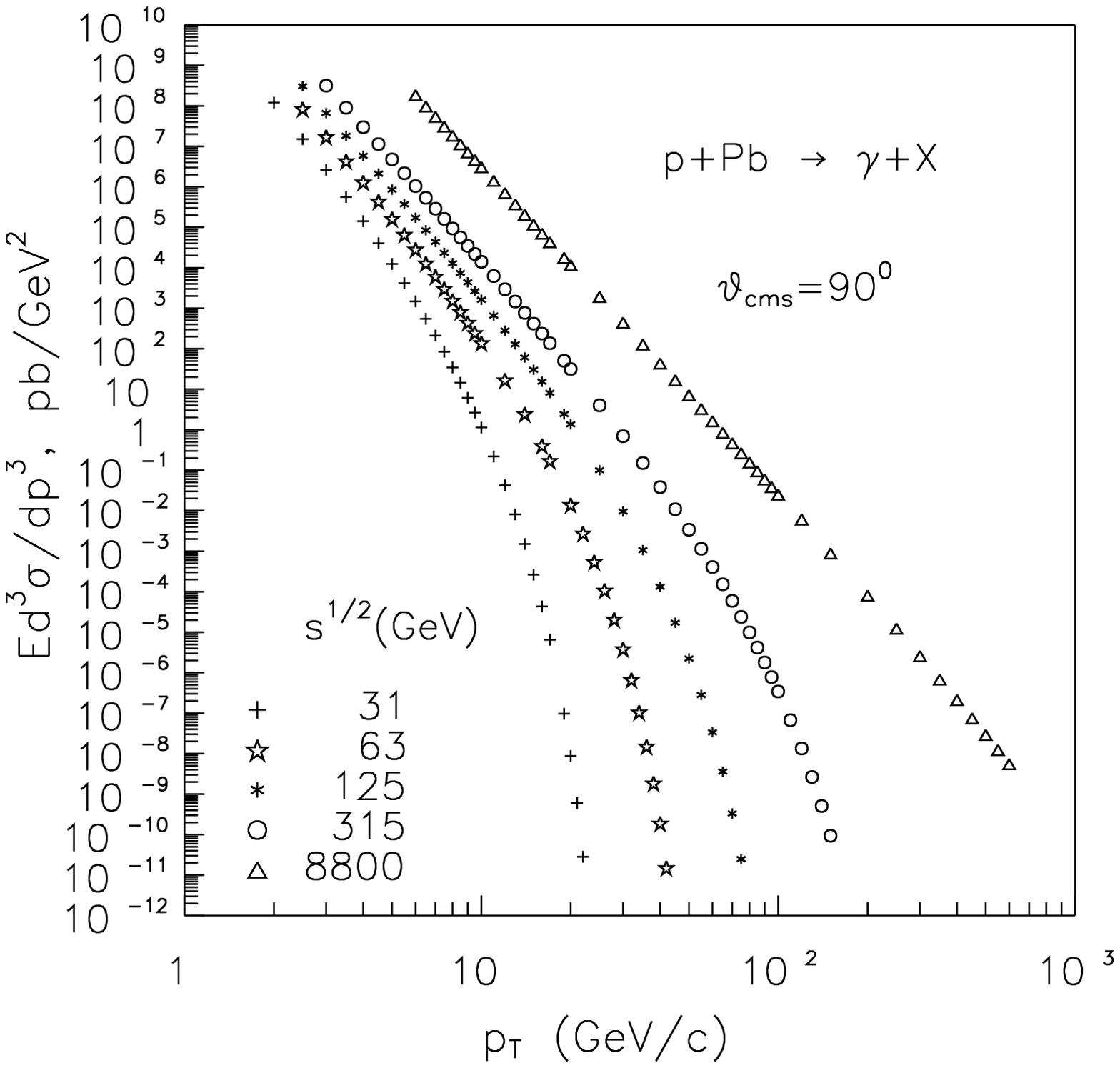}{}}
\hspace*{3cm}
\parbox{5cm}{\epsfxsize=5cm\epsfysize=5cm\epsfbox[95 95 400 400]
{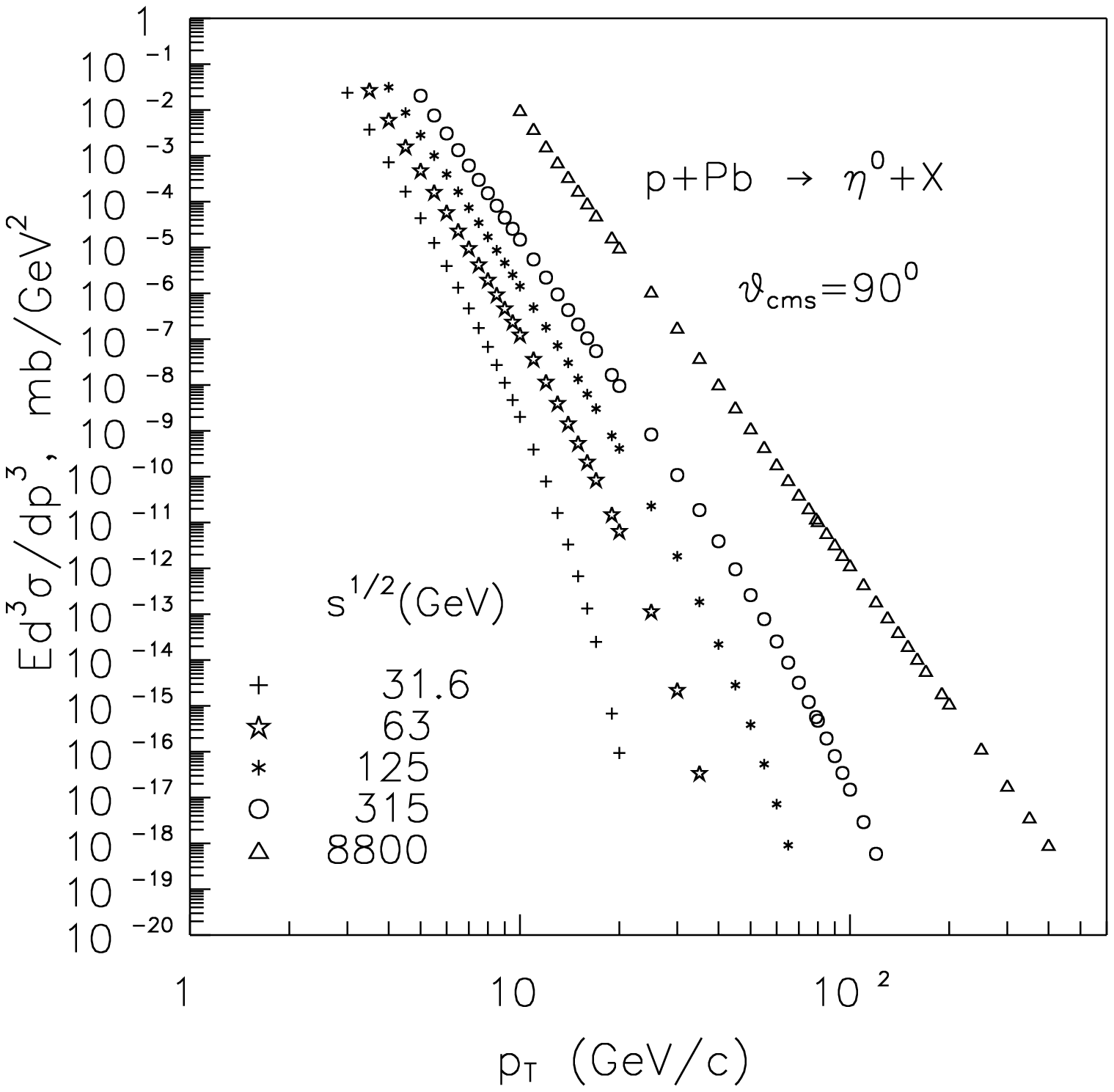}{}}
\vskip -1.cm
\hspace*{0.cm} a) \hspace*{8.cm} b)\\[0.5cm]
\end{center}

{\bf Figure 12.} (a)
 The dependence of
inclusive cross sections of direct photon (a)
and $\eta^0$-meson (b)  production on the transverse momentum $p_T$
in $pPb$ collisions at $\sqrt s = (31-8800)$ GeV. Points ($\triangle,
\circ, *, \star, +$ ) are the calculated results.

\end{document}